\documentclass[twocolumn]{revtex4}

\usepackage{graphicx}

\def\giorno{24/05/2020}

\def\a{\alpha}
\def\b{\beta}
\def\ga{\gamma}
\def\de{\delta}
\def\s{\sigma}
\def\la{\lambda}

\def\^#1{\widehat{#1}}
\def\wt#1{\widetilde{#1}}

\def\beql#1{\begin{equation} \label{#1}}
\def\beq{\begin{equation}}
\def\eeq{\end{equation}}

\def\<{\langle}
\def\>{\rangle}

\def\({\left(}
\def\){\right)}
\def\[{\left[}
\def\]{\right]}

\def\eqref#1{(\ref{#1})}

\def\EOR{\hfill $\odot$}

\def\asymptomaticref{asy1,asy2,asy3,asy4,asy5,asy6,asy7,asy8,asy9,asy10,asy11,asy12,asy13,asy14,asy15,asy16}

\def\fofootnote#1{\footnote{#1}}

\begin{document}

\title{A simple SIR model with a large set of asymptomatic infectives}

\author{Giuseppe Gaeta\thanks{giuseppe.gaeta@unimi.it}}

\affiliation{Dipartimento di Matematica, Universit\`a degli Studi di Milano,
via Saldini 50, I-20133 Milano (Italy) \\ {\rm and} \\ SMRI, 00058 Santa Marinella (Italy) }

\date{\giorno   -- revised \& augmented version}

\begin{abstract}
\noindent
There is increasing evidence that one of the most difficult problems in trying to control the ongoing COVID-19 epidemic is the presence of a large cohort of asymptomatic infectives. We develop a SIR-type model taking into account the presence of asymptomatic, or however undetected, infective, and the substantially long time these spend being infective and not isolated. { We discuss how a SIR-based prediction of the epidemic course based on early data but not taking into account the presence of a large set of asymptomatic infectives would give wrong estimate of very relevant quantities such as the need of hospital beds, the time to the epidemic peak, and the number of people which are left untouched by the first wave and thus in danger in case of a second epidemic wave.} In the second part of the note, we apply our model to the COVID-19 epidemics in Northern Italy. { We obtain a good agreement with epidemiological data; according to the best fit of epidemiological data in terms of this model, only 10\% of infectives in Italy is symptomatic}.
\end{abstract}

\maketitle

\section{Introduction}

There is increasing evidence that one of the main difficulties in trying to control the ongoing COVID-19 epidemic is the presence of a large cohort of \emph{asymptomatic infectives} \cite{\asymptomaticref}. This feature was first noted in the analysis of passengers of evacuation flights from Wuhan, see e.g. \cite{evac}, and on the Crown Princess cruise ship \cite{ship}, but in these circumstances the asymptomatic cases could easily be thought to be just cases which had not yet developed symptoms.

So the first confirmation of this characteristic of COVID was obtained in the large scale study of one of the first infection foci in Italy, that of V\`o Euganeo near Padua; here the whole population (about 3,000 people) of the village were tested twice -- at one week distance -- for the virus, and a significant number of asymptomatic infectives was detected \cite{Crisanti}.

We stress that it is correct to speak of asymptomatic \emph{infectives} and not just \emph{carriers}. In fact, it is by now thought that they are as infective as symptomatic ones in terms of viral charge \cite{asy1,asy13,asy14,asy15} -- and of course potentially even more dangerous as the absence of symptoms leads to a low level of precautions.

One of the consequences of this fact is that the \emph{registered infectives}, those known to the national health systems and thus isolated and monitored, are only a part of the total pool of infectives.

The initial estimates were that registered infectives would be between 1/3 and 1/4 of the actual infectives \cite{ICR}; there have been early claims by the British Government scientific advisers \cite{bbc} that this ratio could be as little as 1/10. In a recent contribution \cite{Li} Li {\it et al.} estimate that only about 1/7 of the infections are detected and can thus be isolated. More recently, other studies have suggested that the fractions of undetected infections could be even higher \cite{Oxf,Imphid}.

This is obviously a very relevant matter, both to understand the epidemic dynamic and to design concrete actions to counter the epidemic spreading.

The goal of this note is two-fold: on the one hand we want to develop a simple general model to take into account the presence of asymptomatic infectives; on the other hand, we want to apply this to the ongoing COVID-19 epidemic, in particular considering the situation in Italy, i.e. in the first European country to be heavily struck by COVID.

One of the problems faced in this situation is that at the start of the epidemic due to a \emph{new} pathogen, one has to estimate the infectivity of this, and more generally the parameters in any model used to describe the epidemic. This will have to be considered in some detail.

{ The estimate of the fraction of asymptomatic infectives is one of the problems encountered in this sense. In the case of COVID in Italy, our estimate -- based on a fit of the epidemiological data by the model, see Sections \ref{sec:numerical} and \ref{sec:measures} below -- is that only 10\% of infectives are symptomatic; this also means that in the early phase of the epidemic, in particular before the relevance of asymptomatic transmission was fully understood and asymptomatic infectives were searched for, only 10\% of the infectives (at best, i.e. assuming that symptoms were properly related to COVID) were registered by the health system.}
\bigskip

The \emph{plan of the paper} is as follows. We start by recalling some basic facts about the well known SIR model (Section \ref{sec:SIR}), and discuss in more detail how this can be fitted against the data available in the first phase of an epidemic (Section \ref{sec:sirfit}).

Beside all the obvious limitations of SIR-type models, the standard SIR -- as recalled above -- does not take into account the special features introduced by the presence of a large set of asymptomatic infectives, i.e. the aspect we want to focus on in this contribution. We will therefore develop a SIR-type model taking into account the presence of asymptomatic infectives, and the substantially long time these spend being infective and not isolated; this is called A-SIR, the A standing indeed for asymptomatic (Section \ref{sec:ASIR}). We also repeat in this context the discussion on how the model parameters can be estimated on the basis of the early stages of the epidemics in that context, see Sections  \ref{sec:ASIRearly} and \ref{sec:asirfit} (we will find that parameters present in both the two models are fitted in the same way from available data).

We will then come, in Section \ref{sec:compare}, to one of the main points of our study, i.e. comparing the different predictions of the standard SIR and of the new A-SIR model for a given set of initial-stage epidemiological data (which are coded by the coefficients of a function fitting them). This comparison is made by means of numerical simulations for realistic values of the parameters, but with no reference yet to any concrete case.

One of our main interests is in understanding how relevant it can be to uncover asymptomatic infectives and promptly isolate them; we then study (numerically) how the dynamics is affected by a reduction of the removal time for this class (Section \ref{sec:ttt}), as could be obtained by a contact tracing and testing campaign \cite{Crisanti}.

The discussion so far considered a \emph{generic} infection (providing long-term immunity to recovered patients) with a large set of asymptomatic infectives. In the second part of the paper, we apply our model to the COVID-19 epidemics in Italy (Section \ref{sec:Italy}). In order to do this we first of all have to estimate the model parameters from data in the initial phase; to this aim we use the data from the first half  of March and determine our best fit of the epidemiological data through a two-step procedure. This first determines relations between the parameters, reducing these to expressions in terms of the the removal times for symptomatic and asymptomatic patients, i.e. on parameters $\b$ and $\eta$ (see below), and this is obtained based on data for the first week; then data for the second week are used to fit these latter parameters. We do of course also compare these fits with the data in the initial phase, see Section \ref{sec:numerical}. The A-SIR model outperforms the standard SIR model in this respect.

We will naturally also consider the subsequent development of the COVID-19 epidemics, but in order to do this we will have to consider the different sets of measures taken on March 8 and March 23 by the Italian Government and based on \emph{social distancing}; these will be taken into account by means of a \emph{reduction of the contact rate} parameter in the model. The amount $r$ of reduction will be a fitted parameter (Section \ref{sec:measures}).

With a suitable choice of $r$, we get a rather good agreement between the dynamic of our model and epidemiological data up to mid-May (i.e. the time of writing of this paper), see Figures \ref{fig:measures} and \ref{fig:RP}.

We will end the paper by a discussion of some relevant general points and by conclusions (Section \ref{sec:conclu}); we anticipate here the main ones:
 \begin{itemize}

 \item[$(i)$] there is a marked difference between the standard SIR dynamics and the dynamics of the A-SIR model, i.e. the one taking into account the presence of a large class of asymptomatic infectives;

 \item[$(ii)$] in the case of the COVID-19 epidemic in Italy, assuming a ratio of symptomatic to total infections of $\xi = 1/10$ yields a good agreement between the model and epidemiological data.

\end{itemize}

\noindent
We stress that our estimate of the ratio $\xi$ is -- as far as we know -- the first one given on the basis of a theoretical model and not just of statistics.
On the other hand, it agrees with the current estimates given by health agencies and physicians.

{ It is maybe useful to also anticipate what are our conclusions about the consequences of the ``marked difference'' mentioned in item $(i)$ above. We find -- in Section \ref{sec:compare} -- that a standard SIR analysis in a
situation where the A-SIR model applies, would make three substantial errors:
$(a)$ The number of (symptomatic) infectives needing Hospital care would be
over-estimated; $(b)$ The time before the epidemic peak – so the time available to prepare the health system to face it – would be over-estimated; $(c)$ The number of people not touched by the epidemic wave, so still in danger if a second wave arises, would be over-estimated. It is rather clear that each of these errors would have substantial practical consequences.}

In an Appendix, we will also discuss how the presence of asymptomatic infectives affects our estimate of the \emph{basic reproduction number} (usually denoted as $R_0$); this might explain why many national health agencies were caught short by the rapid rise in COVID-19 cases.

Our discussion will include a number of small deviations from the central development; these will be given in the form of Remarks. The symbol $\odot$ will signal the end of a Remark.

\section{The SIR model}
\label{sec:SIR}

The classical SIR model for the dynamics of an infective epidemic providing permanent immunity to those who have already been infected and recovered   \cite{KMK,Heth,Murray,Edel,Britton,DH,AM} describes a homogeneous and isolated population of $N$ individuals by partitioning them into three classes: each individual can be either susceptible ($S$), infected and infective ($I$), or removed ($R$) from the epidemic dynamics (that is, either recovered, dead, or isolated). We denote by $S(t)$, $I(t)$ and $R(t)$ the populations of these classes at time $t$; by assumption,
$S(t) +  I(t)  +  R(t)  =  N$ for all $t$.

The model is described by the equations
\begin{eqnarray}
dS/dt &=& - \ \a \, S \, I \nonumber \\
dI/dt &=& \a \, S \, I \ - \ \b \, I \label{eq:SIR} \\
dR/dt &=& \b \,I \ . \nonumber \end{eqnarray}
In the following, the parameter
\beql{eq:gamma} \ga \:= \ \b / \a \eeq will have a special relevance.

This model is well known, but we recall here some of its features both for the sake of completeness and with the purpose of comparing these with those for the new model to be introduced below. Further detail on the SIR model can be found in textbooks \cite{Murray,Edel,Britton,DH,AM}.

We would like to stress a relevant general point. As mentioned above, the SIR model (and more generally SIR type models) assume that we have an isolated population, and that individuals react to the pathogen and interact socially in a homogeneous way. In physicists' language, this is a \emph{mean field} theory, i.e. in common language all individuals characteristics are erased, and real individuals are replaced by an ``ideal'' type corresponding to the average over the population. It goes without saying that these assumptions are not only radical, but also non realistic: individuals in any population differ for age, health state, and contact network. On the other hand, when -- as in a starting epidemic which is overwhelming the sanitary system -- we have scarce data (and available data are not organized in the way they should be in a laboratory experiment) it has the advantage of providing a qualitative description, whose quantitative predictions can be compared with experiment also if fed with the scarce and rough available data.

In other words, we agree that e.g. an epidemic model on  a network \cite{Newman,Anita,Gatto} would be more realistic, but on the one hand we doubt that the available data allow to identify the existing network at this stage\footnote{The health agencies have studied the diffusion of influenza over many years, and this presumably allows to reconstruct some features of the diffusion network (e.g. at department level); but surely the extraordinary alarm caused by COVID emergence has modified all kinds of interpersonal relations and all usual travel patterns among countries or among regions in the same country.}, and on the other hand we would not like to have predictions which depend on the (unknown) features of this network. We will thus be satisfied with working with SIR-type models.

{

\medskip\noindent
{\bf Remark 1.}  Note that according to eqs.\eqref{eq:SIR}, an infected individual is immediately infective. For most infections this is not realistic, of course, but if the delay is substantially smaller that the characteristic removal time $\b^{-1}$, we get a good approximation and still keep to a very simple model with all its advantages for qualitative discussion. The same remark will also apply to the A-SIR model, to be introduced and discussed in Section \ref{sec:ASIR} below.

Note also that the SIR equations \eqref{eq:SIR} stipulate that we have a constant population; in practical terms, this means we are considering an epidemic developing over a short enough timespan, i.e. such that one can disregard deaths and new births. In particular the latter provide new fuel to the susceptible class and thus if the epidemic goes on for a long time related terms should be included in the model; this may led to the presence of an \emph{endemic state} \cite{Murray}. \EOR
}

\subsection{Epidemic dynamics, epidemic peak and total number of infections}

It is immediately apparent that in the SIR model the number of infected will grow as long as \beq S \ > \ \gamma \ ; \eeq thus $\gamma$ is also known as the \emph{epidemic threshold}. The epidemic can develop only if the population is above the epidemic threshold.

{ The ratio
\beql{eq:BRN0} R_0 \ := \ \frac{S_0}{\ga} \eeq is known as the \emph{basic reproduction number} (BRN), or also \emph{basic reproduction rate}, and is an estimate of how many new infections are originated from a single infective in the initial phase of the epidemic \cite{Diek,Diet,Hee}. (In fact, the first equation in \eqref{eq:SIR} says that there are $\a S I (\de t)$ new infections in a time interval $\de t$, and each infective is such for an average time $\b^{-1}$.)}

The parameters $\a$ and $\b$ describe the contact rate and the removal rate; they depend both on the characteristics of the pathogen and on social behavior. For example, a prompt isolation of infected individuals is reflected in raising $\b$, a reduction of social contacts is reflected in lowering $\a$, and both these actions raise the epidemic threshold $\gamma$. If this is raised above the level of the total population $N$, the epidemic stops (which means the number of infected individuals starts to decrease, albeit new individuals will still be infected). The same effect can be obtained by reducing the population $N$ (keeping $\a$ and $\b$ constant), i.e. by partitioning it into non-communicating compartments, each of them with a population below the epidemic threshold.

{
\medskip\noindent
{\bf Remark 2.}
Albeit strictly speaking these predictions only hold within the SIR model, and surely the exact value of the threshold refers to this model only, the mechanism at play is rather general, and similar behaviors are indeed predicted by all kind of epidemic models. \EOR
}
\bigskip

One can easily obtain the relation between $I$ and $S$ by considering the equations governing their evolution in \eqref{eq:SIR} and eliminating $dt$; this provides
\beq dI/dS \ = \ - \, 1 \ + \ \gamma / S \ . \eeq Upon elementary integration this yields (note we always write ``$\log$'' for the natural logarithm)
\beql{eq:I(S)} I \ = \ I_0 \ + \ (S_0 \, - \, S) \ + \ \gamma \, \log(S/S_0) \ ; \eeq with $I_0,S_0$ the initial data for $I(t)$ and $S(t)$; unless there are naturally immune individuals (which is not the case for new infections), $S_0 = N - I_0 \simeq N$.

As we know (see above) that the maximum $I_*$ of $I$ will be reached when $S= \gamma$, we obtain from \eqref{eq:I(S)} an estimate of the level of this maximum; i.e. writing $\gamma = \s N$ (with $\s < 1$) this reads
\beql{eq:I*}  I_* \ = \ \left[ 1 \ - \ \s \ - \ \s \, \log (1/\s ) \right] \ N \ . \eeq
Note that we do \emph{not} have an analytical estimate of the time needed to reach this maximum; see below.

It follows from \eqref{eq:I*} that increasing $\gamma$, even if we do not manage to take it above the population $N$, leads to a reduction of the epidemic peak; if we are sufficiently near to the epidemic threshold, this reduction can be rather relevant also for a relatively moderate reduction of $\a$ and thus increase of $\gamma$.

The formula \eqref{eq:I(S)} also allows to obtain an estimate for another parameter describing the severity of the epidemics, i.e. the total number of individuals $R_\infty$ which are infected over the whole span of the epidemics. In fact, the epidemic is extinct (at an unknown time $t = T_0$)  when $I=0$; the number of susceptibles $S_\infty$ at this stage is provided there by the (lower) root of the equation
$$ I_0 \ + \ (S_0 \, - \, S) \ + \ \gamma \,\log(S/S_0) \ = \ 0 \ ; $$ as noted above $I_0 \simeq 0$, $S_0 \simeq N$, and we can simply look at
\beql{eq:Sinf} (N \, - \, S_\infty ) \ + \ \gamma \, \log(S_\infty / N) \ = \ 0 \ . \eeq
This is a transcendental equation, but it is easily solved numerically if $\ga$ is known. The sought for number of overall infected individuals $R_\infty$ is of course provided by
\beql{eq:Rinf} R_\infty \ = \ N \, - \, S_\infty \ . \eeq

{
\medskip\noindent
{\bf Remark 3.} In the case of a small epidemic, we have $S_\infty/N \simeq 1$, and we can expand the logarithm in a Taylor series; in this way we get $S_\infty= (3 -2 N/\ga) N$. See also Section \ref{sec:KMK} below. \EOR
\bigskip
}

Another key quantity is the \emph{speed} at which the epidemic dynamics develops, and in particular the time $t_*$ at which $I$ reaches its maximum value $I_*$, and the time $t_\infty$ needed for $I$ to get to zero (and $S (t_\infty ) = S_\infty$, of course). In this case one can not get an analytical estimate, but it is possible to describe how this depends on the values of $\a$ and $\b$ for a given population level $N$ and initial conditions $\{ S_0 , I_0 , R_0 \}$.
In fact, the equations \eqref{eq:SIR} are invariant under the scaling
\beql{eq:scale} \a \,\to \, \la \, \a \ , \ \ \b \, \to \, \la \, \b \ , \ \ t \, \to \, \la^{-1} \, t \ . \eeq
(Note that the inverse scaling of $\b$ and $t$ is enforced by the very physical meaning of $\b$, which is the inverse of the characteristic time for the removal of infectives.)

\begin{figure}
\centering
  \includegraphics[width=200pt]{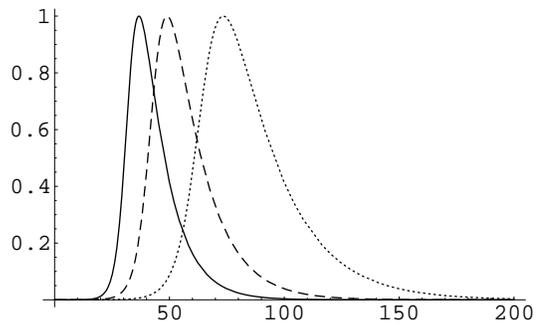}\\
  \caption{{Effect of changing the parameters $\a$ and $\b$ with constant $\ga$ on the SIR dynamics. Numerical solutions to the SIR equations with given initial conditions are computed for different choices of $\a$ and $\b$ with constant ratio $\ga=\b/\a$. In particular we have considered $N=S_0=5*10^7$, $I_0=10^2$, and $\a = \a_0 = 10^{-8}$, $\b = \b_0 = 10^{-1}$ (solid curve); $\a_1 =(3/4) \a_0$, $\b_1 = (3/4) \b_0$ (dashed curve); $\a_2 = (1/2) \a_0$, $\b_2 = (1/2) \b_0$ (dotted curve). The vertical scale is in terms of the maximum $I_*\approx 2.39*10^7$ attained by $I(t)$. This is attained, in the three runs, at times $t_0 \simeq 36.70$, $t_1 = (4/3) t_0 \approx 48.94$ and $t_2 = 2 t_0 \approx 73.41$ respectively.}}\label{fig:SIRTS}
\end{figure}

The meaning of \eqref{eq:scale} is that if we manage to reduce $\a$ by a factor $\la$, even in the case $\b$ is also reduced and thus $\ga$ remains unchanged, the speed of the epidemic dynamics is also reduced by a factor $\la$.
On the other hand, it is clear from the equation for $dS/dt$ in \eqref{eq:SIR} that reducing $\a$ reduces the speed at which new infective appear; if the removal rate $\b$ is unchanged, this will make that $I$ grows slower and reaches a lower level. See Figure \ref{fig:SIRTS} in this regard.

\subsection{Early dynamics}

The SIR equations are nonlinear, and an analytical solution of them turns out to be impossible; they can of course be numerically integrated with any desired precision \emph{if} the initial conditions \emph{and} the value of the parameters are known.

In the case of well known infective agents (e.g. for the influenza  virus) the parameters are known with good precision, and indeed Health Agencies are able to forecast the development of seasonal epidemics with good precision. Unfortunately this is not the case when we face a new virus, as for COVID-19.

Moreover, when we first face a new virus we only know, by definition, the early phase of the dynamics, so parameters should be extracted from such data.\fofootnote{We stress that the parameters depend not only on the pathogen agent, but also on the social structure and organization of the country, e.g. on its population density, restrictive measures, and sanitary system; thus they are different in each country, and data for different countries can not be pooled together to have a wider statistics.} We will thus concentrate on this initial phase, with $I(t)$ and $R(t)$ rather small, and try to obtain approximate analytical expressions for the dynamics; the purpose will be to estimate the parameters $\a$ and $\b$ -- and thus also the epidemic threshold $\ga$ -- in this case.

{
\medskip\noindent
{\bf Remark 4.}
Albeit we do \emph{not} expect, for various reasons, the SIR model to provide a good description of the dynamics when the infection produces a large number of asymptomatic carriers, having an estimate of these parameters will be needed to compare the predictions which one would extract from the standard SIR model in such circumstances with those obtained by the modified model we will consider later on, see Section \ref{sec:ASIR}.
{ As for the reasons to expect the SIR model to perform poorly in the presence of a large set of asymptomatic carriers, these have to do both with the practical implementation and with intrinsic limitations of the model. As for the first type, in practice we estimate the model's parameters by epidemiological data based only on \emph{registered} (thus mostly symptomatic) infectives; if these are only a small part of the infectives, the estimates of the parameters will be grossly different from the true ones. As for the intrinsic limitations of standard SIR in this setting, we will come back to this point at the end of Section \ref{sec:asirmodel}, in Remark 12.}  \EOR
}

\subsection{KMK approximate equations and their exact solution}
\label{sec:KMK}

In the case of ``small epidemics'' there is a way to obtain an analytical expression for the solutions to the SIR equations { or more precisely to the approximate equations valid in the limit of small $R/\ga$}; this is associated to the names of Kermack and McKendrick \cite{KMK}, and we will therefore refer to it as the KMK method. (This is very classical, and is discussed here for the sake of completeness.)

What matters more here, the expression obtained in this way is also an analytical expression holding \emph{in the initial phase} of any epidemics, small or large, i.e. -- as we will discuss in a moment -- until $R(t) \ll \ga$.

Thus such an analytical expression can be compared to \emph{early} epidemiological data and used to estimate the unknown parameters $\a$ and $\b$, and hence the fundamental parameter $\ga$. Once this is done, the model can be studied numerically (or, if we are -- as has to be hoped -- in the favorable situation where $N \simeq \ga$, one can set predictions on the basis of the ``small SIR epidemic'' model) -- recalling of course that the SIR model itself is far too simple to be reliable in a situation where the actions undertaken have heavy consequences on public health -- in order to have some kind of estimate of the length of the epidemics and of other relevant outcomes, such as the numbers $I_*$ and $J_\infty$ considered above.

It should be noted that we do not have full knowledge about the number of infective people at each time; the best we can have is the number of people who are hospitalized or however registered by the health system. Assuming that infective people are immediately isolated, this provides an estimate (actually from below) of $R(t)$. Thus we should be able to compare the predictions for the removed class with epidemiological data, and in order to do this we should focus on  $R(t)$. We stress that his problem was already clear to Kermack and McKendrick \cite{KMK}, see e.g. the discussion in Murray \cite{Murray}, and that we will basically follow their idea albeit with a relevant difference, which will allow for a simpler fit of the data.

Putting together the equations for $S$ and for $R$ in \eqref{eq:SIR}, we have
$dS/dR  = - S/\ga$,
which of course provides
\beql{eq:S(R)} S(R) \ = \ S_0 \ e^{- (R-R_0)/\ga} \ . \eeq
We can proceed similarly with the equations for $I$ and for $R$, getting
$dI/dR = - 1  + S/\ga$, where now $S$ should  be thought of as a function of $R$ through \eqref{eq:S(R)}. Solving this equation we get
\beql{eq:I(R)} I(R) \ = \ I_0
\ + \ S_0 \ \left( 1 \, - \, \exp [ - (R -R_0) / \ga ] \right) \ - \ (R \, - \, R_0 ) \ .  \eeq
We are however interested in the temporal dynamics of the model. In order to do this, we can substitute for $I = I(R)$ using \eqref{eq:I(R)} in the equation for $dR/dt$ in \eqref{eq:SIR}; moreover we will look at the variable
\beq P (t) \ := \ R(t) \ - \ R_0 \ , \eeq which of course satisfies $P (0) = 0$ and $dP/dt = dR/dt$. In this way we have
\beql{eq:I(P)} I(P) \ = \ I_0 \ + \
S_0 \ \left( 1 \, - \, e^{-P/\ga} \right) \ - \ P \ .  \eeq
Plugging now this into the third equation of \eqref{eq:SIR}, we finally get
\beql{eq:P} \frac{dP}{dt} \ = \ \b \ \left[
I_0 \ + \ S_0 \, \left(1 \, - \, e^{-P/\ga} \right) \ - \ P \right] \ . \eeq

This is a transcendental equation and can \emph{not} be solved exactly. However, as long as $P/\ga \ll 1$, i.e. as long as $R(t)$ is well below the epidemic threshold, we can replace the exponential by (a suitable truncation of) its Taylor series expansion.

{
\medskip\noindent
{\bf Remark 5.}
In textbook discussions, it is usually required to consider a \emph{second order}  Taylor expansion; this guarantees that counter-terms preventing the exponential explosion of $R(t)$ (and thus the violation of the condition $R(t) \ll \ga$) are present, and allows to obtain an analytical expression for $R(t)$ { -- solution of the approximate SIR equations for $R/\ga \ll 1$ -- } valid at \emph{all times}.
This is, more precisely, in the form
\beql{eq:RKMK} R(t) \ = \ \frac{\a^2}{S_0} \ \left[ \phi \ + \ k_1 \ \tanh \left[ \frac{k_1 \b}{2} \, t \ - \ k_2 \right] \ \right] \ , \eeq where we have written $\phi := (S_0/\ga -1)$ and  $k_1$ and $k_2$ are explicitly given by
\beq
k_1 = \sqrt{ \phi^2 + 2 (S_0/\ga^2) (N - S_0)} ; \
k_2  = k_1^{-1} \, \mathrm{arctanh} (\phi ) \ . \eeq
As we assume there is no natural immunity, we can take $S_0 \approx N$, obtaining $k_1 \approx \phi$ and hence slightly simpler complete expressions. \EOR

\medskip\noindent
{\bf Remark 6.}
In particular, in this case the maximum of $R'(t)$ -- and hence of $I(t)$, see \eqref{eq:SIR} -- is obtained at time
$$ t_* \ = \ \frac{2 \ \mathrm{arctanh} (\phi )}{\b \ \phi^2} \ ; $$
as our result holds for the ``small epidemics'', $\phi$ is small and we can write
$$ \wt{t}_* \ \simeq \ \frac{2}{\b \,\phi} \ + \ \frac23 \frac{\phi}{\b} \ . $$
Note that $t_*$ is therefore rapidly decreasing with $\phi$ (for small $\phi$). On the other hand, looking back at \eqref{eq:I*}, and noticing that in terms of the notation used there $\s = 1/(1+\phi)$, we obtain immediately that $I_*$ grows with $\phi$. \EOR

\medskip\noindent
{\bf Remark 7.}
This also means that if one would be able to tune the parameters $\a$ and $\b$ (and hence $\phi$) there would be a contrast between trying to have a low $I_*$ and hence a small $\phi$, and trying not to have the epidemic running for too long -- which can be devastating on social and economic grounds. If, on the other way, the priority from the temporal point of view is on slowing down the epidemic, e.g. to have the time to prepare the health system facing the peak, a small $\phi$ should be pursued. \EOR

}

\bigskip

\subsection{Small time solution of the KMK equations}
\label{sec:KMKshort}

We have seen in Section \ref{sec:KMK} that a full solution of the KMK model is equivalent to a full solution of eq. \eqref{eq:P}, and if we have a small epidemic then the solution is well approximated -- \emph{at all times} -- by \eqref{eq:RKMK}.

Here we are less ambitious: when the epidemic is just starting, we can in any case only fit the initial phase of the epidemic, which shows an exponential increase of $R(t)$, and correspondingly we can expand the exponential in \eqref{eq:P} at \emph{first} order in $P/\ga$. This yields the equation
\beql{eq:SerP} \frac{dP}{dt} \ = \ \b \
\left[ I_0 \ + \ \left( \frac{S_0}{\ga} \, - \, 1 \right) \ P \right] \ ,
\eeq
with initial condition $P(0) =0$. This is immediately solved to give
\beq P(t) \ = \ I_0 \ \frac{\exp[ \b \, (S_0/\ga - 1) \, t ] \ - \ 1}{(S_0 / \ga \, - \, 1 )} \ . \eeq
Introducing the parameter, which we stress is now \emph{not} assumed to be small,
\beql{eq:phi} \phi \ := \ \frac{S_0}{\ga} \ - \ 1 \ , \eeq  the above is more simply written as
\beql{eq:P(t)} P(t) \ = \ \frac{I_0}{\phi} \ \left[e^{\b \, \phi \, t} \ - \  1 \right] \ , \eeq
and finally we get
\beql{eq:R(t)} R(t) \ = \ R_0 \ + \ \frac{I_0}{\phi} \ \left[e^{\b \, \phi \, t} \ - \  1 \right] \ . \eeq
As expected this -- at difference with \eqref{eq:RKMK} -- is not saturating but just expanding exponentially, and thus cannot be valid for all times, but only for $t$ sufficiently small (even for small $\phi$).

The expression \eqref{eq:R(t)} can then be expanded in series to give the small $t$ expression of the solution, which can be fitted against experimental data thus determining (some of) the parameters, see Section \ref{sec:sirfit} below.

{

\medskip\noindent
{\bf Remark 8.}
It is relevant for the following of our discussion to note that the solution \eqref{eq:R(t)} can be obtained also in a different way, i.e. noticing that in the initial phase of the epidemic the number of susceptibles vary very little and can thus be considered as constant, $S(t) \simeq S_0$. Within his approximation, and writing again $\phi = (S_0/\ga -1)$, the SIR equations reduce to
\beq \cases{ dI/dt \ = \ \b \, \phi \, I & \cr dR/dt \ = \ \b \, I & ; \cr} \eeq
this is a linear system, and it is promptly solved to yield indeed \eqref{eq:R(t)}.

Note this approach is actually simpler than the one followed above (so the reader may wonder why we have not taken this immediately), but on the one hand the procedure given above is along the lines of the tradition of SIR analysis, and on the other hand  having seen that derivation gives us more confidence that a rough approach as this one provides the same results as a more refined one; this will be of use dealing with more complex models, where the Kermack-McKendrick approach can not be followed, see Section \ref{sec:ASIRearly} below. \EOR

}

\subsection{Fitting the SIR parameters}
\label{sec:sirfit}

Note that the solution \eqref{eq:R(t)} depends on three parameters, i.e. $\b$, $I_0$ and $\phi$, which in turn depends on the known number $S_0 \simeq N$ and on the unknown parameter $\ga$. None of the parameters $\{ \b , \phi , I_0 \}$ is known, but $\b$ can be directly estimated on the basis of medical data as it corresponds to the inverse of the typical removal time (for trivial infections, this corresponds to the time of healing; in the case of COVID it is the time from infection to isolation), and similarly once we fix a time $t=0$ the number $I_0$ can be estimated \emph{a posteriori} looking at epidemiological data for the next few days and depending on our estimate of $\b$.

In order to estimate the parameters on the basis of the measurements of $R$, we can work either on $R$ itself, or on its logarithm. That is, we have two alternative -- but obviously equivalent -- ways to proceed.

\bigskip\noindent{\bf (a)} {\it Working on the time series for $R(t)$}. \par\noindent We fit the time series around $t_0$ by
\beql{eq:Rsergen} R(t) \ = \ r_0 \ + \ r_1 \, t \ + \ \frac12 \, r_2 t^2 \ ; \eeq Having these coefficients $r_k$, we can compare with the series expansion for $R(t)$ given by \eqref{eq:R(t)}, which is just
\beql{eq:RserSIR} R(t) \ = \ R_0 \ + \ (\b \, I_0) \ t \ + \
\frac12 \, \left( \b^2 \, I_0 \, \phi \right) \ t^2
\ . \eeq
We obtain easily that -- using also the definition of $\phi$ \eqref{eq:phi} -- our parameters and the associated parameter $\ga$ are given by
\begin{eqnarray} R_0 &=& r_0 \ , \ \ I_0 \ = \ \frac{r_1}{\b} \ , \ \ \phi \ = \ \frac{r_2}{r_1 \ \b} \ ; \nonumber \\
\ga &=& \frac{\b S_0 r_1}{\b r_1 + r_2} \ . \label{eq:linfitsir} \end{eqnarray}

\bigskip\noindent{\bf (b)} {\it Working on the time series for $\log[R(t)]$.} \par\noindent As $R(t)$ grows -- in the early phase -- in a substantially (but not exactly, see above) exponential way, one usually deals with data in logarithmic form; that is one has a fit for $\log [R(t)]$, say of the form
\beql{eq:LRsergen} \log[R(t)] \ = \ A \ + \ B \, t \ + \ \frac12 \, C \ t^2 \ . \eeq
Comparing with the series expansion of $\log [R(t)]$ for $R$ as in \eqref{eq:R(t)}, i.e.
\begin{eqnarray} \log [ R(t)]  &=& \log (R_0 ) \ + \ \b \ \frac{I_0}{R_0} \ t \nonumber \\ & & \ - \ \frac12 \ \left( \frac{\b \, I_0}{R_0} \right)^2 \, \left( 1 - \phi \frac{R_0}{I_0} \right) \ t^2 \ , \label{eq:LRsersir} \end{eqnarray}
we obtain that the $I_0$, $\phi$ and $\ga$ parameters can be estimated as
\begin{eqnarray} R_0 &=& e^A \ , \ \ I_0 \ = \ \frac{B}{\b} \ e^A \ , \ \ \phi \ = \ \frac{B^2 \ + \ 2 \, C}{\b \ B} \ ; \nonumber \\ \ga &=& \frac{\b S_0 B}{\b B + B^2 + 2 C} \ . \label{eq:logfitsir} \end{eqnarray}
\bigskip

We stress again that our estimates for the parameters depend on our estimate for the leading parameter $\b$.

This concludes our discussion of the standard SIR model.

\section{A model with asymptomatic infectives}
\label{sec:ASIR}

It may happen to have an epidemic such that a rather large fraction of infected people are actually asymptomatic, but still fully infective; this appears to be the case for COVID-19.\fofootnote{Albeit in the first stage it was believed that asymptomatic were not fully infectives, now they are thought to be so  \cite{\asymptomaticref}.}

A little reflection shows that the presence of a large population of asymptomatic infectives, or however of infectives which show only very mild symptoms, easily thought not to be related with the concerned infective agent, changes the dynamics in two -- contrasting -- ways:

\begin{itemize}

\item On the one hand, asymptomatic infectives are a formidable vehicle of contagion, as they have no reason to take special precautions, and get in contact with a number of people which themselves do not take the due precautions (which would be taken in the case of an individual with evident symptoms);

\item On the other hand, assuming once the infection is ceased the infected (including of course asymptomatic ones) have acquired permanent immunity, they contribute to group immunity reached once the population of susceptibles falls below the epidemic threshold.

\end{itemize}

\noindent
We are thus going to study how the SIR dynamics is altered by the presence of a large class of asymptomatic infectives.

{

\medskip\noindent
{\bf Remark 9.} An obvious but important Remark is in order here. If we find out that known infectives are only a fraction $\xi < 1$ of the total infectives, this means that on the one hand the mortality rate (number of deceased over number of infected) is actually smaller by the same factor. On the other hand, the total number of infected persons is increased by a factor $\xi^{-1}$, so that it looks more difficult to stop the spreading of the epidemics, and the final number of infected will be quite large.

In this respect, one should however remember that the total number of casualties does not depend only on the total number of individuals with symptoms but also on the number of patients needing Intensive Care (for the COVID epidemic in China this was estimated at 20 \% of hospitalized patients \cite{CDC,WHOCDC}; in Italy this fraction reached 10\% at peak time, albeit in the most affected regions the number was higher \cite{WHOrep,PCrep,MinSal}) and on the availability of such care; from this point of view, slowing down the pace of the epidemics can substantially lower the death toll even if the total number of affected individuals remains the same. \EOR
}

\subsection{The A-SIR model}
\label{sec:asirmodel}

We will formulate a very simplified model of SIR type, where infective people are either symptomatic or asymptomatic. A more refined subdivision of their state would be more realistic, but the discussion of this simple case will suffice to show how to proceed in a more general setting.\footnote{A notable example of these more detailed models for COVID is provided by \cite{pavia}. In our opinion keeping to a simple setting will help focusing on the essential mechanism at hand, albeit when facing a concrete epidemic the health authorities would probably prefer to follow a less generic model -- if available, and if the available data are sufficient to reliably fit all the many parameters appearing in it.}

In our model we still assume permanent immunity of individuals who have been infected and recovered, and constant population. We also assume that -- as in the classical SIR -- infected individuals are immediately infective (see Remark 1 in this regard). We will have susceptibles $S(t)$ in a unique class, but two classes of infected and infective people: symptomatic $I (t)$ and asymptomatic $J (t)$; and similarly two classes of removed people: formerly symptomatic removed $R(t)$ and formerly asymptomatic removed (mostly passing unnoticed through the infection) $U(t)$. Symptomatic infectives are removed by the epidemic dynamics through isolation (in hospital or at home) at a removal rate $\b$ (thus with typical delay $\b^{-1}$, while asymptomatic people are removed from the epidemic dynamics mostly through spontaneous recovery, at a recovery rate $\eta \ll \beta$, thus after a typical time $\eta^{-1} \gg \b^{-1}$; detecting asymptomatic infectives leads to their isolation before healing, and thus to a reduction in $\eta^{-1}$.

We assume that both classes of infected people are infective in the same way, and that an individual who gets infected belongs with probability $\xi$ to the class $I$ and with probability $(1-\xi)$ to the class $J$.

Our model, which we will call A-SIR (Asymptomatic-SIR) will then be
\begin{eqnarray}
dS/dt &=& - \, \a \ S \, (I + J ) \nonumber \\
dI/dt &=& \a \, \xi \, S \, (I +J) \ - \ \b \, I \nonumber \\
dJ/dt &=& \a \, (1-\xi) \, S \, (I +J) \ - \ \eta \, J \label{eq:ASIR} \\
dR/dt &=& \b \, I \nonumber \\
dU/dt &=& \eta \, J \ . \nonumber \end{eqnarray}
Note that the last two equations amount to an integral, i.e. are solved by
\begin{eqnarray} R(t) &=& R_0 \ + \ \b \ \int_0^t I(\tau) \ d \tau \ , \nonumber \\
U(t) &=& U_0 \ + \ \eta \ \int_0^t J(\tau) \ d \tau \ ; \label{eq:RU}
\end{eqnarray} thus they can be considered as ``reconstruction equations'', and we will focus on the first three equations in \eqref{eq:ASIR}. Moreover, the total population $N = S+I+J+R+U$ is constant.

{
\medskip\noindent
{\bf Remark { 10}.} The fact that asymptomatic infectives are as infective as symptomatic ones is not at all obvious. In the case of COVID-19 it appears we are exactly in this situation \cite{asy1,asy13,asy14,asy15}; in any case this will be our working hypothesis. \EOR
}

{
\medskip\noindent
{\bf Remark { 11}.} An alternative writing for the equations \eqref{eq:ASIR} is also possible, and in some case convenient. We denote by
\beql{eq:K} K(t) \ = \ I(t) \ + \ J(t) \eeq the total number of infectives, and by
\beql{eq:x} x(t) \ = \ I(t)/K(t) \eeq the fraction of symptomatic infectives; obviously the fraction of asymptomatic ones will be $y(t) = J(t)/K(t) = 1 - x(t)$.

By standard algebra the (first three) equations \eqref{eq:ASIR} for the A-SIR dynamics are then rewritten in terms of these variables as
\begin{eqnarray}
dS/dt &=& - \, \a \, S \, K \nonumber \\
dK/dt &=& \a \, S \, K \ - \ \[ \b \, x \ + \ \eta \, (1- x) \] \, K \label{eq:asirX} \\
dx/dt &=& \a \, S \, (\xi - x) \ - \ (\b -\eta) \, (1 - x) \ . \nonumber \end{eqnarray}
This writing also stresses that albeit $\xi$ is a constant, the fraction $x$ of symptomatic infectives is a \emph{dynamical variable}. \EOR
}

{ {
\medskip\noindent
{\bf Remark 12.} We can now better discuss, in the light of the previous Remark 11, what are the intrinsic reasons which make that we expect the standard SIR model to perform poorly in describing an epidemic with a large number of asymptomatic infectives (see Remark 4 above). As already recalled, one would expect $\b$ to correspond to the inverse of the average removal time for registered infectives; however, once the presence of asymptomatic -- and in particular undetected -- infectives is ascertained, this could be corrected by accepting as $\b$ a weighted average $B$ of the removal times for symptomatic and asymptomatic (that is, of $\b$ and $\eta$ in the A-SIR model). But as we have just seen in Remark 11, the fraction $x$ of symptomatic infectives is a dynamical variable, hence the parameter $B$ should vary with time depending on the internal dynamics of the system (which is described by A-SIR, not by SIR), and hence we would be outside the proper SIR framework. \EOR } }

\subsection{A-SIR dynamics}

Some general considerations on eqs.\eqref{eq:ASIR} can be made immediately. First of all, we note that $I(t)$ will increase as far as the condition
$ \a  \xi S  (I +J) > \b  I $ is satisfied; that is, as far as
\beql{eq:ga1} S \ > \ \ga_1 \ := \ \frac{1}{\xi} \ \frac{\b}{\a} \ \frac{I}{I+J} \ = \ \frac{x}{\xi} \ \frac{\b}{\a} \ . \eeq
Thus the epidemic threshold (for symptomatic patients) $\ga_1$ depends both on the fixed parameters $\xi,\a,\b$ \emph{and} on the variable ratio $x(t)$ of symptomatic infectives over total infectives, see \eqref{eq:x}.

Similarly, the number of asymptomatic infectives $J(t)$ will grow as far as
$ \a  (1-\xi)  S  (I +J) > \eta  J  $ is satisfied, i.e. as far as \beql{eq:ga2} S \ > \ \ga_2 \ := \ \frac{1}{1-\xi} \ \frac{\eta}{\a} \ \frac{J}{I+J} \ = \ \frac{1-x}{1-\xi} \ \frac{\eta}{\a} \ . \eeq
Again the epidemic threshold (for asymptomatic patients) $\ga_2$ depends both on the fixed parameters $\xi,\a,\eta$ \emph{and} on the variable ratio $y(t) = 1 - x(t)$ of a\-sympto\-ma\-tic -- and thus ``hidden'' --  infectives over total infectives.

Note that
$$ \frac{\ga_1}{\ga_2} \ = \ \left( \frac{1-\xi}{\xi} \right) \ \left( \frac{\b}{\eta} \right) \ \left( \frac{I}{J} \right) \ = \ \left( \frac{1-\xi}{\xi} \right) \ \left( \frac{\b}{\eta} \right) \ \left( \frac{x}{1-x} \right)\ . $$

As we expect on the one hand to have $\xi < 1/2$ and $\beta > \eta$, but on the other hand $I < J$ and hence $x < 1-x$, we cannot claim there is a definite ordering between $\ga_1$ and $\ga_2$; this means that we will have situations where $I$ declines and $J$ is still growing, but the opposite is also possible.

We expect that in the very first phase --when the different removal times have not yet shown their effects -- we have $ J \simeq [(1-\xi)/\xi] I$ and hence
$(1-x) \simeq [(1-\xi)/\xi] x$; under this condition, we get $$ \frac{\ga_1}{\ga_2} \  \simeq \ \frac{\b}{\eta} \ > \ 1 \ . $$

{
\medskip\noindent
{\bf Remark 13.} In the case of COVID-19, it is known that the incubation time is about $5.1$ days; assuming that symptomatic infection is promptly recognized and swiftly treated, epidemiological and clinical data suggest the approximate values (note that asymptomatic removal time $\eta^{-1}$ includes both the incubation time and the healing time)
$\b^{-1} \simeq 5 - 7$, $\eta^{-1} \simeq 14 - 21$ for the removal and recovery rates; the value of $\xi$ is more controversial, as mentioned in the Introduction, but it is presently believed that $\xi \simeq 1/10$. We will come back to these matters when dealing specifically with COVID, but wanted to give immediately an idea of what ``realistic'' values can be for the parameters appearing the A-SIR model. \EOR
}

\subsection{Early dynamics}
\label{sec:ASIRearly}

It is quite clear that we can not go through the Kermack-McKendrick procedure to obtain approximate equations valid in the case of ``small epidemics'', not even through the simplified (first rather than second order) procedure we have used above.
We can however go through the even simpler approach mentioned in Remark 8 (and which we have seen there produces the same results as the KMK procedure).

With $S(t) \simeq S_0$, the equations \eqref{eq:ASIR} reduce to a \emph{linear} system of four equations with constant coefficients, or more precisely to a ``master'' system of two equations
\begin{eqnarray}
\frac{dI}{dt} &=& (\a \, \xi \, S_0 \, - \, \b ) \, I \ + \ (\a \, \xi \, S_0 ) \, J  \label{eq:smI} \\
\frac{dJ}{dt} &=& [\a \, (1-\xi) \, S_0] \, I \ + \ [\a \, (1-\xi) \, S_0 \, - \, \eta ] J \label{eq:smJ}  \end{eqnarray}
plus the two direct integrations \eqref{eq:RU}.\fofootnote{Note  that if we use the alternative formulation \eqref{eq:asirX}, setting $S \simeq S_0$ still leaves us with nonlinear equations.}

As for the two equations, \eqref{eq:smI} and \eqref{eq:smJ},
we can get their solution in explicit form by means of some standard algebra; they are slightly involved when written in fully explicit form, and we do not report them here.

With these, we can compute $R(t)$ and $U(t)$; their explicit expressions are also quite involved, and again we do not report them here.

{

\medskip\noindent
{\bf Remark 14.} As already remarked -- and as well known from textbooks, see e.g. Murray \cite{Murray} -- we can only access $R(t)$ from epidemiological data.In our case, this is the number of symptomatic patients which have been registered and thus isolated, i.e. in particular removed from the epidemic dynamic. As for $U(t)$, this time series is basically unattainable: some of the asymptomatic infectives will be discovered and registered, but many -- and probably most -- of them will go unnoticed. Thus we should base our considerations \emph{only} on $R(t)$. Unfortunately, the data published by health agencies and by WHO do not distinguish between symptomatic and asymptomatic patients. \EOR
}

\subsection{Fitting the parameters}
\label{sec:asirfit}

We can now proceed as in Section \ref{sec:sirfit}, i.e. series expand $R(t)$ in order to fit the parameters. From the explicit expression of $R(t)$ we get
\begin{eqnarray} R(t) & \simeq & R_0 \ + \ \b \, I_0 \ t \nonumber \\
&+& \frac12 \, \b \left[ \a (I_0 + J_0) S_0 \xi \, - \, \b I_0 \right] \ t^2 \ ; \label{eq:Rser} \\
\log [R(t)] & \simeq & \log (R_0) \ + \ \b \, \frac{I_0}{R_0} \ t \nonumber \\
 &+&\frac12 \ \frac{\b [\a (I_0 + J_0) R_0 S_0 \xi \, - \ \b I_0 (I_0 + R_0) ]}{R_0^2} \ t^2 \ . \label{eq:Rlogser}  \end{eqnarray}

Comparing these with the generic form\fofootnote{Which we repeat here for convenience of the reader: $R(t) \simeq r_0 + r_1 t  + (1/2) r_2 t^2$,
$\log[R(t)] \simeq A + B  t + (1/2) C  t^2$.} of the fits \eqref{eq:Rsergen} and \eqref{eq:LRsergen},
we can express the parameters $I_0$ and $\ga = \b/\a$. Note that we can not express both $\ga$ and $J_0$ with the same fitting, as both of them only appear in the coefficient of the quadratic term. Note also that in this context $\ga$ is \emph{not} any more the epidemic threshold, as discussed in Section \ref{sec:asirmodel}; the time-varying epidemic threshold $\ga_1 (t) = [x(t)/\xi] \ga$ is however expressed in terms of $\ga$, so that it still makes sense to fit it.

Actually, since new infected are with probability $\xi$ in the class $I$ and with probability $(1-\xi)$ in the class $J$, it is natural to set as initial condition
\beql{eq:j20} J_0 \ = \ \left( \frac{1-\xi}{\xi} \right) \ I_0 \ ; \eeq
with this assumption, we have
\beq \ga_1 (t_0) \ = \ \ga \ . \eeq

{

\medskip\noindent
{\bf Remark 15.} It should be noted that actually if we want to fit $\ga$ we need to have some estimate on $J_0$ (while $I_0$ can be fitted from first order coefficient in the series for $R(t)$ or $\log[R(t)]$); to this aim we will use consistently \eqref{eq:j20}. \EOR
}
\bigskip

In particular, using the fit of $R$ we get (through this assumption)
\beql{eq:linasirfit} R_0 \ = \ r_0 \ , \ \ I_0 \ = \ \frac{r_1}{\b} \ , \ \ \ga \ = \ \frac{\b \, r_1 \, S_0}{(\b \, r_1 \ + \ r_2)} \ . \eeq

Using instead the fit of $\log[R(t)]$, and again the assumption \eqref{eq:j20}, we get
\begin{eqnarray} R_0 &=& e^A \ , \ \ I_0 \ = \ \frac{B \, e^A}{\b} \ ; \nonumber \\ \ga  &=&  \frac{\b \, S_0 \, B}{\b \, B \ + \ B^2 \ + \ 2 \, C} \ . \label{eq:logasirfit} \end{eqnarray}

Once the parameters are estimated, the nonlinear equations \eqref{eq:ASIR} can be solved numerically.

It is immediate to check that the expressions \eqref{eq:linasirfit} and \eqref{eq:logasirfit} -- which we recall were obtained under the assumption \eqref{eq:j20} for $J_0$ -- are \emph{exactly} the same as for the SIR model; see \eqref{eq:linfitsir} and \eqref{eq:logfitsir}.

\section{Comparing SIR and A-SIR dynamics}
\label{sec:compare}

Summarizing our discussion so far, we have considered both the standard SIR model and a variant of it, the A-SIR model, taking into account the presence of large set of asymptomatic infectives. We have discussed in particular how the parameters for the models can be estimated on the basis of the time series for $R(t)$ in the early stage of the epidemic.

We want now to discuss how the prediction extracted from a given time series for $R(t)$ in the beginning of an epidemic differ if these are analyzed using the SIR or the A-SIR models.

As discussed above, we are able to extract only limited analytical information from the nonlinear SIR equations \eqref{eq:SIR}, and no relevant analytical predictions at all from the nonlinear A-SIR equations \eqref{eq:ASIR}. Thus the only way to compare the predictions of the latter model with those of a standard SIR model (or variations on it, such as the SEIR model \cite{Murray}), and hence  see how the presence of a large class of asymptomatic infectives affects the dynamics within a SIR-type framework, is at present by running numerical simulations, i.e. numerically integrate the SIR and the A-SIR equations for \emph{coherent} sets of parameters.

Here by ``coherent'' we will mean ``extracted from the same time series for $R(t)$ in the early phase of the epidemics''.

In practice, in view of the discussion above and depending on our choice to use the \eqref{eq:linasirfit} or the \eqref{eq:logasirfit} fit, this means ``with given $\{r_0,r_1,r_2;\b\}$ or given $\{ A,B,C;\b \}$ coefficients''.

In the following, we will analyze the concrete epidemiological data for the COVID-19 epidemics in Italy, but here we would like to compare the SIR and the A-SIR predictions in a less concrete case, so that one can focus on general features rather than having a concrete case (with all the intricacies of real cases, see below) in mind.

We will thus choose an arbitrary (but realistic) set of parameters, which should be thought of as being extracted from the time series for the early phase of the epidemic. We consider a total population $N = 10^7$, $S_0 = N$, and parameters
\begin{eqnarray} & & A \ = \ 5 \ , \ \ B \ = \ 0.1 \ , \ \ C \ = \ -0.002 \ ; \nonumber \\ & & \b \ = \  \frac{1}{10} \ , \ \ \eta \ = \ \frac{1}{20} \ ; \ \ \xi \ = \ \frac{1}{10} \ . \label{eq:parcomp} \end{eqnarray}
Moreover, we will assume that asymptomatic infectives (in the A-SIR modeling) are totally undetected; that is, we compare predictions which would be made by a mathematician using the SIR model and totally unaware of the existence of asymptomatic infectives with those made by a mathematician using the A-SIR model and aware of the relevance of asymptomatic infectives.

\begin{figure}
\centering
\begin{tabular}{cc}
  \includegraphics[width=100pt]{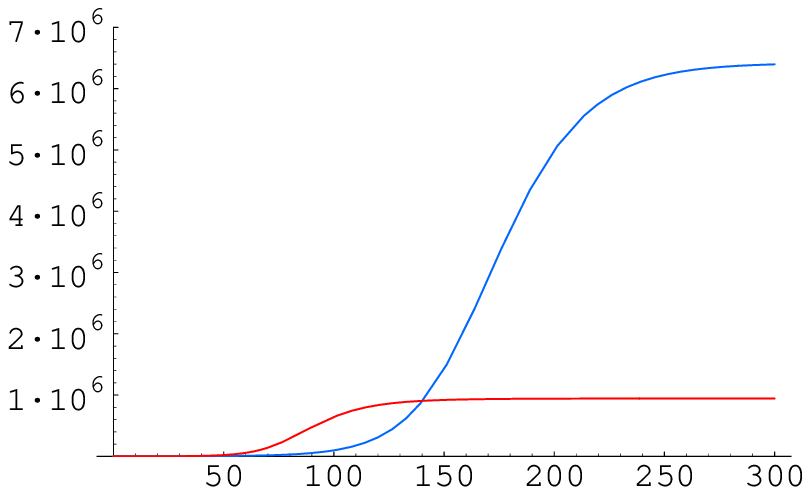} &
  \includegraphics[width=100pt]{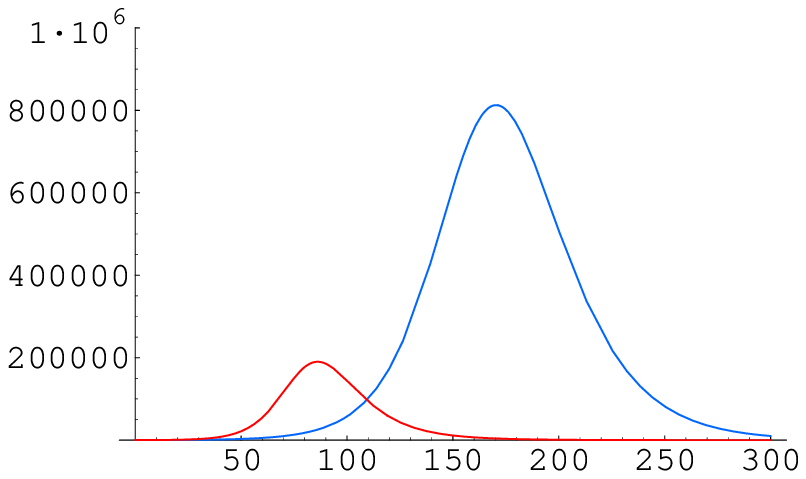} \\
  $R(t)$ & $I(t)$ \end{tabular}
  \caption{{  The predictions for the cumulative number of removed $R(t)$ and for the (detected) infectives $I(t)$ provided by the SIR model (blue curves) and by the A-SIR model (red curves) for a total population $N=10^7$ and the parameters given in eq.\eqref{eq:parcomp}.}}\label{fig:compRJ}
\end{figure}

In Fig.\ref{fig:compRJ} we compare the predictions issued by the SIR and the A-SIR models for the cumulative number of (detected) removed, i.e. for $R(t)$, and for the number of (detected, hence symptomatic) infectives $I(t)$.

Note that not only the SIR predicts a higher infective peak, but it also expects it to occur at a later time; moreover for the initial phase, in particular after the A-SIR peak, the prediction of the number of infectives issued by the SIR model are \emph{lower} than those issued by the A-SIR model. Note also that the SIR model expects a much greater part of the population to go through symptomatic infection, as seen from the $R(t)$ plot.

\begin{figure}
\centering
\begin{tabular}{cc}
  \includegraphics[width=100pt]{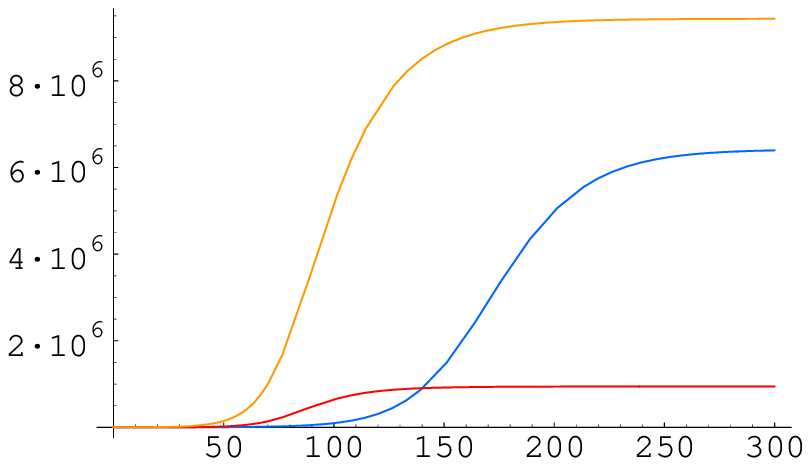} &
  \includegraphics[width=100pt]{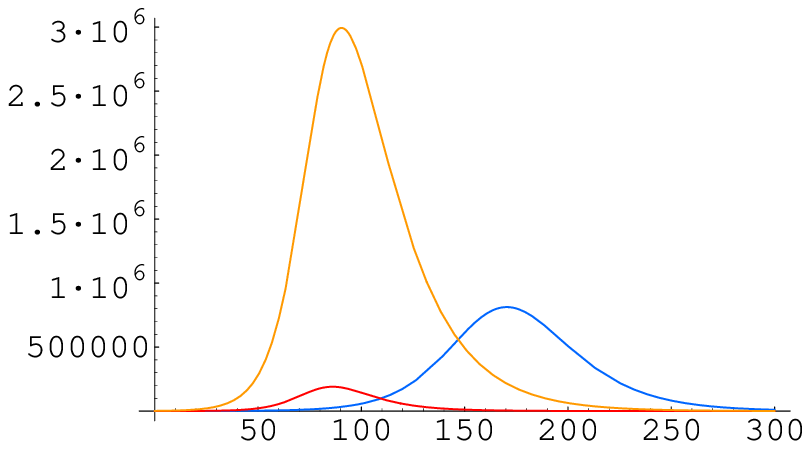} \\
  $R(t)$ and $[R(t)+U(t)]$ & $I(t)$ and $[I(t)+J(t)]$ \end{tabular}
  \caption{{  The predictions for the cumulative number of removed $R(t)$ and for the (detected) infectives $I(t)$ provided by the SIR model (blue curves) and those by the A-SIR model (red curves) together with the predictions for $R(t)+U(t)$ and for $I(t) +J(t)$ provided by the A-SIR model (orange curves) for a total population $N=10^7$ and the parameters given in eq.\eqref{eq:parcomp}.}}\label{fig:comp_all}
\end{figure}

These discrepancies are reduced, and somehow reversed, if in the A-SIR modeling we consider \emph{both} symptomatic and asymptomatic infected. These graphs are plotted in Fig.\ref{fig:comp_all}.

\begin{figure}
\centering
\begin{tabular}{cc}
  \includegraphics[width=100pt]{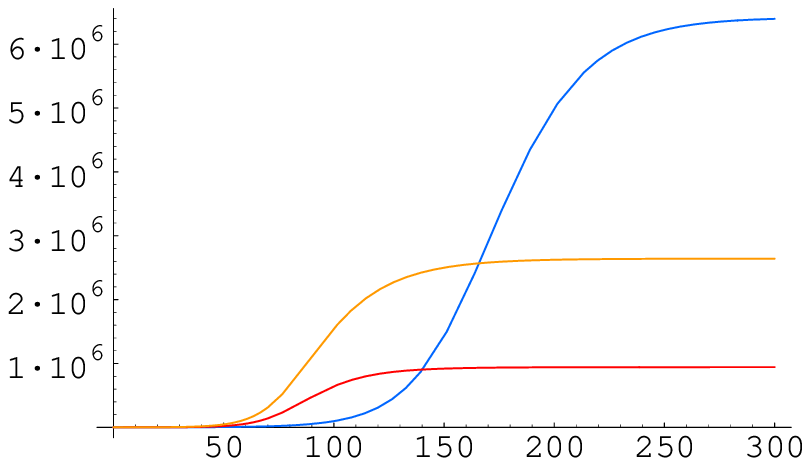} &
  \includegraphics[width=100pt]{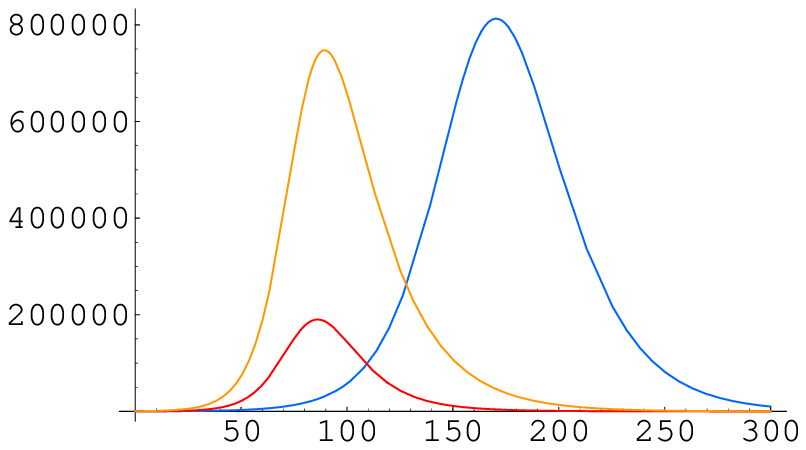} \\
  $R(t)$ and $[R(t)+ \s U(t)]$ & $I(t)$ and $[I(t)+ \s  J(t)]$ \end{tabular}
  \caption{{  The predictions for the cumulative number of removed $R(t)$ and for the (detected) infectives $I(t)$ provided by the SIR model (blue curves) and those by the A-SIR model (red curves) together with the predictions for detected removed  $R(t)+ \s U(t)$ and detected infectives $I(t) + \s J(t)$ provided by the A-SIR model (orange curves) if a fraction $\s = 1/5$ of  asymptomatic infectives is detected, for a total population $N=10^7$ and the parameters given in eq.\eqref{eq:parcomp}.}}\label{fig:comp_part}
\end{figure}

In a realistic situation one should expect that at least apart of the asymptomatic infectives are anyway detected (e.g. being tested due to contacts with known infectives). We have thus plotted the equivalent of Fig.\ref{fig:comp_all} in Fig.\ref{fig:comp_part}, supposing that a fraction (1/5) of asymptomatic infectives is detected. Note that in this case the height of the epidemic peak is about the same as for the SIR model, but it occurs at an earlier time, while the number of detected removed when the epidemic is over is still substantially lower (about half) of the one predicted by the SIR model.

In other words, a modeler issuing his/her advice based on a SIR analysis in a situation where the A-SIR model applies, would make three substantial errors:

\begin{itemize}

\item The number of (symptomatic) infectives needing Hospital care would be over-estimated;
\item The time before the epidemic peak -- so the time available to prepare the health system to face it -- would be over-estimated;
\item The number of people not touched by the epidemic wave, so still in danger if a second wave arises, would be over-estimated.

\end{itemize}

It is rather clear that all of these errors would have substantial consequences.
We stress that one could naively think that the presence of asymptomatic infections is essentially of help, in that a large part of the population getting in contact with the virus will not need any medical help. This is true, but on the other hand we have seen that it will also make that the curve raises more sharply than it would be expected on the basis of the SIR model (see also the discussion in Appendix A), and that in all the first phase of the epidemic -- until the true epidemic peak and also for a period after this -- the number of symptomatic infectives is higher than what would be foreseen in the basis of the SIR model.

We conclude that it is absolutely essential to take into account the presence of asymptomatic infectives.

\section{Tracing, Testing, Treating}
\label{sec:ttt}

In the previous Section \ref{sec:compare} we have seen how the presence of asymptomatic infectives changes the predictions which would be done on the basis of the standard SIR model. In the present Section we want to consider a problem which naturally arises when consider an infection with a large number of asymptomatic infectives; that is, how the epidemiological dynamics is changed by an extensive campaign of \emph{testing} (maybe \emph{tracing} contacts of known infectives), followed of curse by \emph{treating} -- which in the case of asymptomatic means essentially just isolating the infectives so they they do not spread the infection.\footnote{The title of the Section refers to the fortunate slogan proposed in the COVID-19 context. Much before this slogan was proposed, this approached was followed (in a low-tech mode) by the Padua group to tackle the COVID-19 epidemics in the field in V\`o Euganeo and then in Veneto, with remarkable success \cite{Crisanti}.}

To this aim, we will consider again -- obviously just in the context of the A-SIR model -- the situation seen in the previous Section, but will take into account the effects of such a ``triple T'' action by a reduction of the average removal time $\tau = \eta^{-1}$ for asymptomatic infectives. (A more detailed study is given in our related paper \cite{Gavsb}.)

\begin{figure}
\centering
\begin{tabular}{cc}
  \includegraphics[width=100pt]{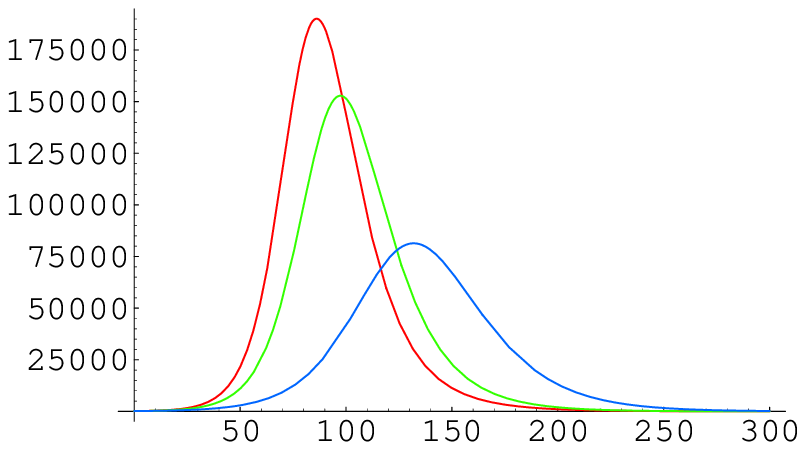} &
  \includegraphics[width=100pt]{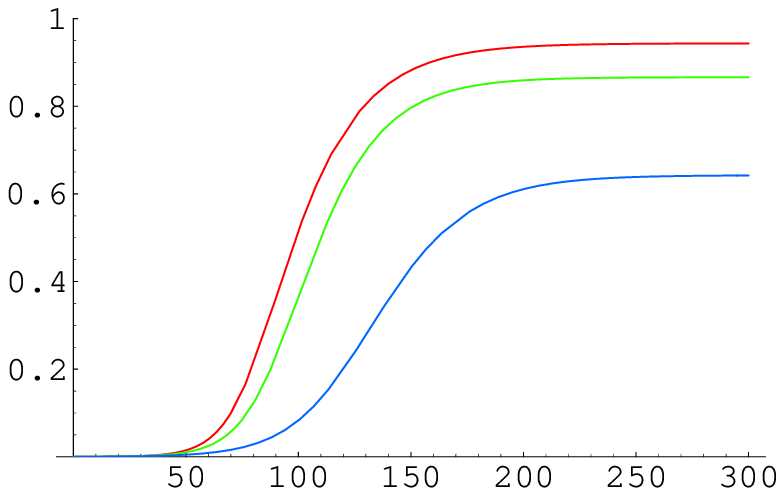} \\
  $I(t)$ & $R(t)+U(t)$ \end{tabular}
  \caption{{  The predictions for the number of symptomatic infectives $I(t)$ and for the total number of removed (and thus in the end immune) patients $R(t)+U(t)$ provided by the A-SIR model for a total population $N=10^7$ and the parameters given in eq.\eqref{eq:parcomp}, with $\eta = 1/20$ (red curves), $\eta = 1/15$ (green curves) and $\eta = 1/10 = \b$ (blue curves).}}\label{fig:ttt}
\end{figure}

We will thus run numerical simulations with the same parameters (except for $\eta$), total population and initial conditions as in Section \ref{sec:compare} above, and choose three different values for $\eta \le \b$, i.e. $\b= 1/20, 1/15,1/10$. The result of these simulations is displayed in Fig.\ref{fig:ttt}.

We see from these that a ``Tracing, Testing, Treating'' campaign can reduce substantially the epidemic peak\fofootnote{More precisely, Fig.\ref{fig:ttt} displays a reduction of the \emph{symptomatic} epidemic peak -- which is the relevant one in terms of charge on the health system -- but the plot for the total infected would be analogous, except of course for a different scale.}; on the other hand this leads to a slowing down of the epidemics, and to a smaller number of total infected individuals, i.e. to a smaller immune population at the end of the epidemic wave -- and hence smaller resistance in the case of a second wave.

This concludes our general discussion. In the following, we will apply our formalism to the analysis of the COVID-19 epidemics in Italy.

\section{The COVID-19 epidemics in Italy}
\label{sec:Italy}

In the previous Sections, we have conducted our analysis in general terms. This referred to a generic infection with a large number of asymptomatic infectives, but of course the motivation for it was provided by the ongoing COVID-19 pandemic.

We will study in particular the situation in Italy; as well known, this was the first European country heavily struck by COVID, and data for it are widely available. Moreover, when COVID landed in Europe there was already an alert, so that transient phenomena due to late recognition of the problem -- which were unavoidable in Eastern Asia -- were of a much smaller size. Last but not least, in this case we are confident to have all the relevant information.

\subsection{Epidemiological data}

The data for the cumulative number of registered infected communicated by the Italian Health System \cite{PCrep} -- and also available through the WHO reports \cite{WHOrep} -- are given in Table \ref{tab:I}. There we give data for February 21 to May 15, albeit we will for the moment only use those for the first half of March; later data will be of use in the following. (We prefer not to use the data for the first few days, as too many spurious effects an affect these.)

One should note, in this respect, that the first cases in Italy (apart from sporadic and promptly isolated cases of tourists) were discovered on February 21. The public awareness campaign started immediately -- and actually the public was already alert, e.g. it was impossible to find face masks since some weeks -- and the first local mild restrictive measures were taken a few days later (February 24), and more restrictive measures involving the most affected areas\fofootnote{Due to a leak of information, a number of people fled from the most affected area before the prohibition to do so went into effect; this has most probably pushed the spreading of the infection in different regions, but luckily the Southern regions, which were expected to be heavily affected by this, in the end were only marginally touched.} were taken on March 1. A more stringent set of measures went into effect for the whole nation on March 8, and still more strict measures on March 22.

Thus the epidemics developed with varying parameters. Moreover, as the incubation time for COVID ranges from 2 to 10 days, with a mean time of 5.1 days \cite{ICR},  there is a notable delay in the effect of any measure. Note that most countries have a quarantine length of 14 days; thus we expect that any measure will show their effect with a delay of one to two weeks.

\begin{table}
\centering
\begin{tabular}{||l||r|r|r|r|r|r||}
\hline
day & Feb 21 & Feb 22 & Feb 23 & Feb 24 & Feb 25 \\
R & 20 & 77 & 146 & 229 & 322 \\
\hline
day & Feb 26 & Feb 27 & Feb 28 & Feb 29 & Mar 1 \\
R & 400 & 650 & 888 & 1128 & 1694 \\
\hline
day & Mar 2 & Mar 3 & Mar 4 & Mar 5 & Mar 6  \\
R & 1835 & 2502 & 3089 & 3858 & 4636 \\
\hline
day & Mar 7 & Mar 8 & Mar 9 & Mar 10 & Mar 11  \\
R & 5883 & 7375 & 9172 & 10149 & 12462 \\
\hline
day & Mar 12 & Mar 13 & Mar 14 & Mar 15 & Mar 16  \\
R & 15113 & 17660 & 21157 & 24747 & 27980 \\
\hline
day & Mar 17 & Mar 18 & Mar 19 & Mar 20 & Mar 21  \\
R & 31506 & 35713 & 41035 & 47021 & 53578 \\
\hline
day & Mar 22 & Mar 23 & Mar 24 & Mar 25 & Mar 26  \\
R & 59138  & 63927 & 69176 & 74386 & 80539 \\
\hline
day & Mar 27 & Mar 28 & Mar 29 & Mar 30 & Mar 31  \\
R & 86498 & 92472 & 97689 & 101739 & 105792 \\
\hline
day & Apr 1 & Apr 2 & Apr 3 & Apr 4 & Apr 5  \\
R & 110574 & 115242 & 119827 & 124632 & 128948 \\
\hline
day & Apr 6 & Apr 7 & Apr 8 & Apr 9 & Apr 10  \\
R & 132547 & 135586 & 139422 & 143626 & 147577 \\
\hline
day & Apr 11 & Apr 12 & Apr 13 & Apr 14 & Apr 15  \\
R & 152271 & 156363 & 159516 & 162488 & 165155 \\
\hline
day & Apr 16 & Apr 17 & Apr 18 & Apr 19 & Apr 20  \\
R & 168941 & 172434 & 175925 & 178972 & 181228 \\
\hline
day & Apr 21 & Apr 22 & Apr 23 & Apr 24 & Apr 25  \\
R & 183957 & 187327 & 189973 & 192994 & 195351 \\
\hline
day & Apr 26 & Apr 27 & Apr 28 & Apr 29 & Apr 30  \\
R & 197675 & 199414 & 201505 & 203591 & 205463 \\
\hline
day & May 1 & May 2 & May 3 & May 4 & May 5  \\
R & 207428 & 209328 & 210717 & 211938 & 213013 \\
\hline
day & May 6 & May 7 & May 8 & May 9 & May 10  \\
R & 214457 & 215858 & 217185 & 218268 & 219070 \\
\hline
day & May 11 & May 12 & May 13 & May 14 & May 15  \\
R & 219814 & 221216 & 222104 & 223096 & 223885 \\
\hline
\end{tabular}
    \caption{{  Cumulative number $R(t)$ of COVID-19 registered infect in Italy  \cite{PCrep,MinSal}. In our fits for the early phase of the epidemic, $t=t_0$ corresponds to March 5 and $(t_i,t_f)$ to the period March 1 through March 10.}}\label{tab:I}
\end{table}

\subsection{Fit of the data}
\label{sec:ItaFit}

For our fitting in the early phase, we will consider the data of the period March 1 through March 10, denoted in the following as $t_i$ and $t_f$ respectively; this leaves us some later days to compare the functions obtained through the fit with subsequent evolution, before the effect of the first set of measure can show up (March 15).

\begin{figure}
\centering
  \includegraphics[width=100pt]{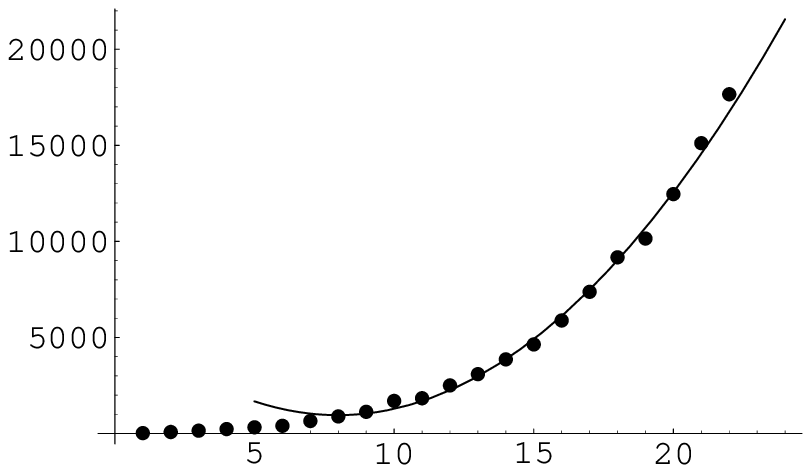} \ \
  \includegraphics[width=100pt]{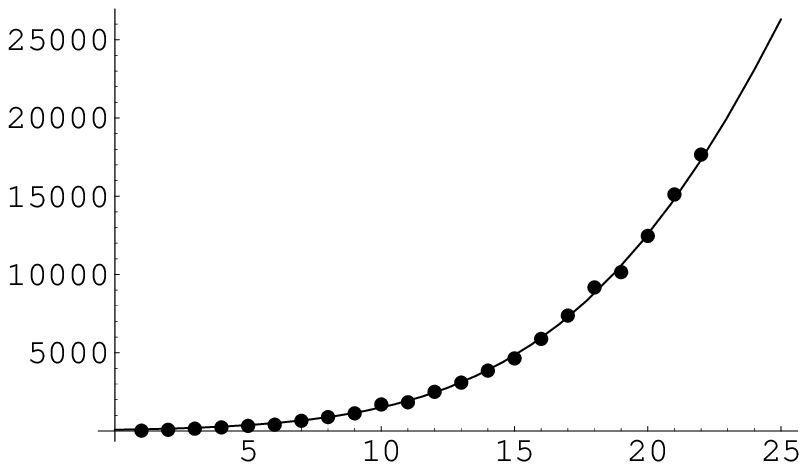} \\
  \caption{{  Data for $R(t)$ in the COVID epidemics in Northern Italy from February 24 to March 13, with fits obtained using data for March 1 through March 10. Time is measured in days, with day zero being February 20.  Left: polynomial (quadratic) fit \eqref{eq:Rsergen}; Right: corrected exponential fit \eqref{eq:LRsergen}.}}\label{fig:plotsdata}
\end{figure}

The best direct fit of $R(t)$ through a quadratic function, see \eqref{eq:Rsergen},
is obtained with the constants
\beq r_0 \ = \ 3862.32 \ , \ \ r_1 \ = \ 966.54 \ , \ \ r_2 \ = \ 80.35 \ . \eeq
The fit is reasonably good in the considered time interval $(t_i,t_f)$, but fails completely for $t < t_i$ (see Figure \ref{fig:plotsdata}, where we use data from February 24 on) and is rather poor for $t > t_f$. This is not surprising, as we know that $R(t)$ is, in this early phase, growing through a slightly corrected exponential law, see \eqref{eq:R(t)}.\fofootnote{Note however that here the fit has not the goal to provide an analytical description of $R(t)$ for a larger interval of time, but only to estimate some parameters for the nonlinear SIR or A-SIR equations.}

Let us then look at the fit of $\log[R(t)]$ as in \eqref{eq:LRsergen}.
In this case the best fit is obtained with the constants
\beq A \ = \ 8.26648 \ , \ \ B \ = \  0.221083 \ , \ \ C \ = \ - 0.00430354 \ . \eeq
In this case the fit is very good not only within $(t_i,t_f)$ but also outside it, at least for the period until March 15. We will thus work only with this (exponential) fit.

We will consider these values for the coefficients $\{r_0,r_1,r_2 \}$ or for the coefficients $\{A,B,C\}$ as experimental measurements.

We can now use the formulas obtained before, both for the SIR and the A-SIR model, to estimate the parameters of these models in terms of these fits following the discussion in Sections \ref{sec:sirfit} and \ref{sec:asirfit}.

\subsection{Estimate of SIR and A-SIR parameters}

We have remarked in Section \ref{sec:asirfit} that the SIR and A-SIR models (in the latter case, under the assumption \eqref{eq:j20} for $J_0$) yield exactly the same values for the $I_0$ and $\ga$ parameters. Now we want to estimate these values for the data given in Section \ref{sec:ItaFit}; this amounts to a direct application of formulas \eqref{eq:linasirfit} and \eqref{eq:logasirfit} -- or equivalently \eqref{eq:linfitsir} and \eqref{eq:logfitsir}, as already remarked.

As already remarked, these formulas depend on the estimated value of the parameter $\b$. The values obtained using the direct fit of $R$ are tabulated for different values of $\b$ in the upper part of Table \ref{tab:II}.

We can also proceed by using the fit of $\log[R(t)]$; the values obtained in this way are tabulated for different values of $\b$ in the lower part of Table \ref{tab:II}.

\begin{table}
\centering
\begin{tabular}{|r||c||l|l|l|l|l|l|l|l||}
\hline
QF & $\b$ & 1/3 & 1/4 & 1/5 & 1/6 & 1/7 & 1/8 & 1/9 & 1/10 \\
\hline
 & $I_0$ & 2900 & 3866 & 4833 & 5799 & 6766 & 7732 & 8699 & 9665\\
 & $\phi$ & 0.25 & 0.33 & 0.42 & 0.50 & 0.58 & 0.67 & 0.75 &  0.83 \\
 & $S_0/\ga$ & 1.25 & 1.33 & 1.42 & 1.50 & 1.58 & 1.67 & 1.75 & 1.83 \\
\hline
\hline
EF & $\b$ & 1/3 & 1/4 & 1/5 & 1/6 & 1/7 & 1/8 & 1/9 & 1/10 \\
\hline
 & $I_0$ & 2581 & 3441 & 4301 & 5162 & 6022 & 6882 & 7743 & 8602 \\
 & $\phi$ & 0.55 & 0.73 & 0.91 & 1.09 & 1.28 & 1.46 & 1.64 & 1.82 \\
 & $S_0/\ga$ & 1.55 & 1.73 & 1.91 & 2.09 & 2.28 & 2.46 & 2.64 & 2.82 \\
\hline \end{tabular}
\caption{{  Upper part (quadratic fit -- QF): Parameters for the SIR and A-SIR models obtained through the SIR quadratic local fit of $R(t)$; see \eqref{eq:linfitsir}, \eqref{eq:linasirfit}. Lower part (exponential fit -- EF): Parameters for the SIR and A-SIR models obtained through the SIR modified exponential local fit of $R(t)$; see \eqref{eq:logfitsir}, \eqref{eq:logasirfit}.}}\label{tab:II}
\end{table}

We remind that the delay time $\delta$ from infection to arise of symptoms is estimated to be around $\delta \simeq 5.2$ \cite{ICR}; thus albeit we have tabulated several options for $\b$, values greater than $1/5$ are not realistic, at least in the absence of an aggressive contact tracing and tracking campaign.

On the other hand, albeit we expect $\b^{-1}$ to be around one week for \emph{symptomatic} patients, it should be recalled that in the standard SIR framework this parameter refers to the average removal time for \emph{all} patients, symptomatic and asymptomatic. Thus the presence of asymptomatic ones could make $\b^{-1}$ to be substantially larger. We will leave this remark aside for the time being; see the discussion in the Appendix, in particular Remark A1.

In all cases, $S_0/\ga$ is quite far from one and $\phi$ from zero, so one can not rely on the ``small epidemic'' formulas \cite{Murray} discussed in Section \ref{sec:KMK}. We will thus resort to numerical integration, see next Section.

This concludes our estimation of the SIR or A-SIR parameters from the epidemiological data referring to the early phase of the epidemic.

\section{Numerical simulations. Timescale of the epidemic}
\label{sec:numerical}

We have now determined the parameters for both the SIR and the A-SIR model in the very early phase of the epidemic; this determination depends actually on the value of $\b$, so from now on we will consider $\b$ as the only parameter to be fitted.

We can now run numerical simulations with the SIR and the A-SIR equations; we can vary $\b$ (and thus implicitly also the other parameters, maintaining the relation between these and $\b$) and fit $\b$ according to the agreement between the outcome of the numerical simulations and the epidemiological data.

In all of our simulations, day one is February 21, so the fitting period $(t_i,t_f)$ corresponding to the first decade of March used to determine the relation between $\b$ and the other parameters is centered around day 14; but this does not determine $\b $ itself. We will now examine the fit of data in the first period outside this window in order to determine $\b$.

\subsection{General study}
\label{sec:detail}

Note that while the discussion of the previous Section \ref{sec:Italy} is complete for what concerns the SIR parameters, in the A-SIR equation we also need to introduce the removal rate for asymptomatic individuals, i.e. $\eta$; this is related to the time length $\delta = \eta^{-1}$ of their infective period, which is equal to the incubation time plus the spontaneous healing time\footnote{More precisely, this would be the time until the viral load becomes too low to infect other individuals. From this point of view, different attitudes exist in different countries, also concerning isolation requirements. E.g., in Italy patients who tested positive but with little or light symptoms and are thus isolated at their home, are required to remain isolated until testing negative in \emph{two} swab tests, while in Germany and other countries, they are automatically authorized to leave home after 14 days.}. While the former is around $\de \simeq 5$, the latter is generally considered to be around 14 days, albeit we know that for hospitalized patients (which however are by definition symptomatic) this may be longer.
Thus, on these medical grounds, we expect $\eta^{-1}$ to be between 14 and 21 days.

\subsubsection{Determining $\b$}

We ran a number of simulations, both for the SIR and the A-SIR dynamics, with varying $\b$ and -- for the A-SIR model -- with varying $\eta$; the parameter $\b$ was varied in the range $(1/10-1/5)$, while $\eta$ in the range $(1/25,1/10)$.

\begin{figure}
\centering
\begin{tabular}{|c|c|}
\hline
  \includegraphics[width=100pt]{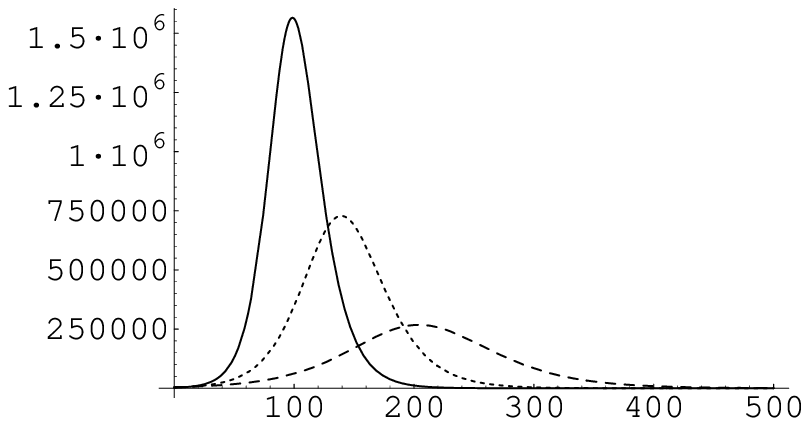} &
  \includegraphics[width=100pt]{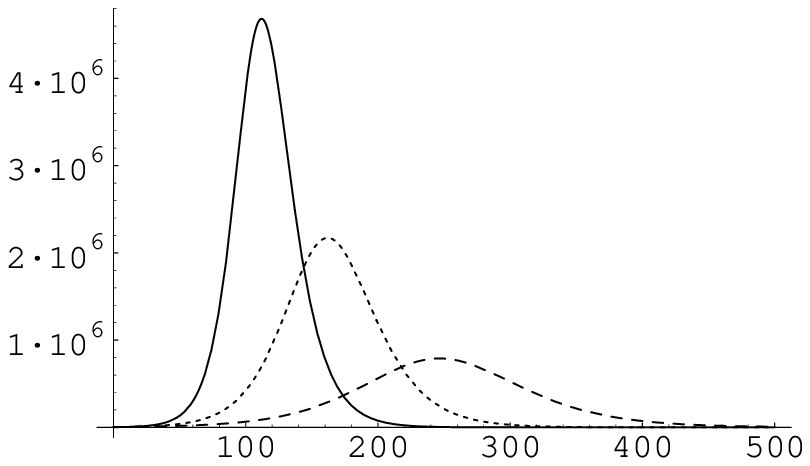} \\
$S_0 = 2*10^7, \ \b = 1/7 $ & $S_0 = 6*10^7, \ \b = 1/7 $ \\
\hline
    \includegraphics[width=100pt]{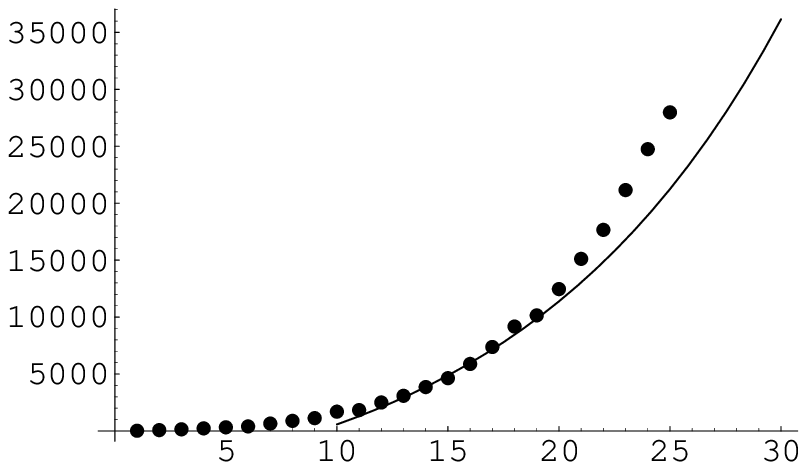} &
    \includegraphics[width=100pt]{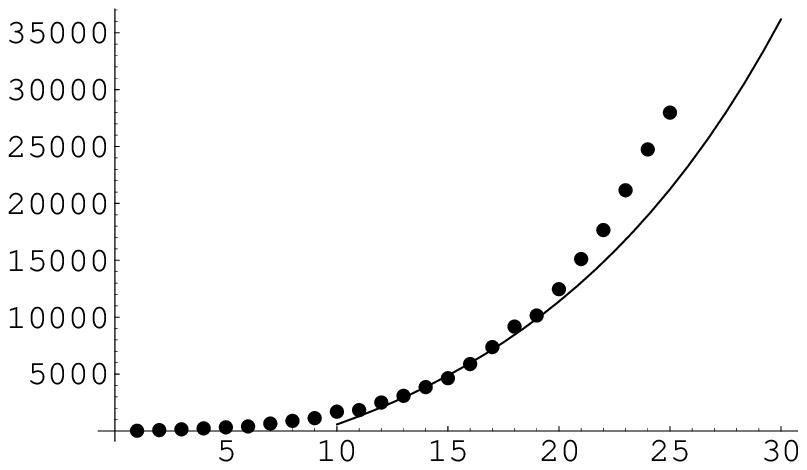} \\
$S_0 = 2*10^7, \ \b = 1/7 $ & $S_0 = 6*10^7, \ \b = 1/7 $ \\
\hline
\end{tabular}
\caption{Upper row: numerical solution of the SIR equations for $\b=1/7$ and total population $S_0= 2*10^7$ (left) and $S_0=6*10^7$ (right), corresponding to the overall population of the three Northern Italy regions most affected by COVID and to the population of all of Italy respectively, using for the parameters $I_0$ and $\ga$ the fit of eqs. \eqref{eq:linasirfit}, \eqref{eq:logasirfit} on the basis of the data for March 1 through March 10, see Table I. The plots of $I(t)$ -- where $t$ is measured in days -- are shown for: $r=1$ (solid curve), $r=0.85$ (dotted curve) and $r=0.75$ (dashed curve). Lower row: plot of the data for the COVID epidemics in Italy for March 1 through March 17 (hence outside the fitting region) versus the numerical integration of the SIR model (with $r=1$). The SIR model is not properly describing the dynamics.}\label{fig:plotsirbest}
\end{figure}

\begin{figure}
\centering
\begin{tabular}{|c|c|}
\hline
  \includegraphics[width=100pt]{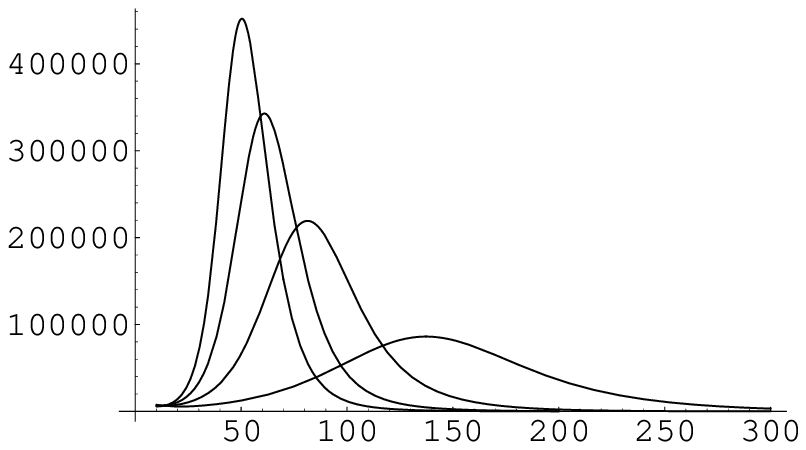} &
  \includegraphics[width=100pt]{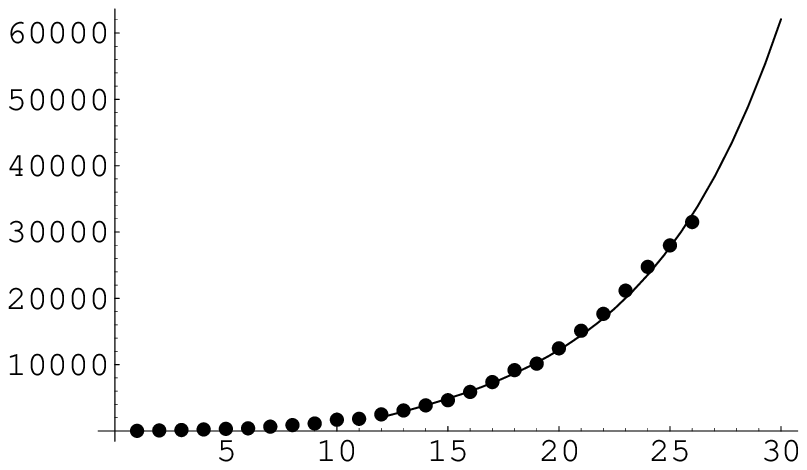} \\
  \hline
\end{tabular}
  \caption{Numerical solution of the A-SIR equations for $S_0 = 2*10^7$, $\b = 1/7$, $\eta = 1/21$, $\xi = 1/10$, using for the parameters $I_0$ and $\ga$ the fit of eqs. \eqref{eq:linasirfit}, \eqref{eq:logasirfit}  on the basis of the data of Table I for the period 1-10 March. Left: plots of $I(t)$ -- where $t$ is measured in days -- are shown for: $r=1$, $r=0.8$, $r=0.6$ and $r=0.4$ (the curves for higher $r$ are those with higher peak). Right: Plot of the data for the COVID epidemics in Italy versus the numerical integration of the A-SIR model (with $r=1$); plotted data go until March 17.} \label{fig:plotasirbest}
\end{figure}

The best results in terms of agreement with epidemiological data for the period March 1 through March 15 were obtained for
$$ \b \ = \ \frac{1}{7} \ , \ \ \eta \ = \ \frac{1}{21} \ . $$
The plot for these values of the parameters are  displayed in Figs.\ref{fig:plotsirbest} and \ref{fig:plotasirbest}; note that for the SIR case the fit is rather poor.

In fact, it should be stressed that the situation is quite different in the cases of SIR and of A-SIR dynamics. In particular, the SIR dynamics does not fit the data for the week after the fitting window, while the A-SIR dynamics fits these quite well.

We stress that the fit $\b \simeq 1/7$ is coherent with medical data: in fact, as already recalled, the average delay from infection to the first symptoms is about 5 days, and there should be some further delay for recognizing these symptoms -- which in the first phase can be rather trivial, such as cold, cough, or light fever -- as due to COVID and leading to hospitalization or anyway to isolation.\fofootnote{This even disregarding the fact that at the peak of the epidemic the medical system was completely overwhelmed in several Departments -- especially in Lombardia -- and action was delayed.}

{

\medskip\noindent
{\bf Remark 16.} A relevant point should be made here. In the first phase of the epidemic, and in particular in the period covered by Figs.\ref{fig:plotsirbest} and \ref{fig:plotasirbest}, the COVID diffusion was essentially limited to Northern Italy, and in particular to the three regions of Lombardia, Veneto and Emilia-Romagna; the total population of Italy is about 60 millions, while that of the three mentioned regions totals about 20 millions. This is why simulations with both $S_0 = 2*10^7$ and $S_0 = 6*10^7$ are shown in these Figures. Actually, the comparison with the epidemiological data \emph{in the early stage} should be better done with the simulation with $S_0=2*10^7$ (we are giving the other one for the SIR model only to show that even considering the overall population we do not get a reasonable fit). \EOR
}
\bigskip

\subsubsection{Timescale of the epidemic}

In Section \ref{sec:compare} above we have discussed the different timescale of an epidemic with given initial time series predicted by the SIR and by the A-SIR models. It is natural to wonder how long the COVID epidemic will last according to these.

This is not a well posed question, because there are restrictive measures being taken (or relaxed at a later stage, when the situation improves) which reduce the contact rate and thus the spread of the epidemic.

So, these simulations can at their best show what would be the behavior (of the system described by the SIR or A-SIR equations, which do not necessarily describe correctly the COVID epidemics) \emph{with constant coefficients}. On the other hand, they can give an idea of what should be expected in case of no action.

More generally, and in line with our general discussion of Sections \ref{sec:compare} and \ref{sec:ttt}, they show how the dynamics predicted by the SIR and the A-SIR models with coherent sets of parameters differ from each other in general (albeit using the concrete COVID parameters to fix ideas).

It should be stressed in this context that the containment measures based on \emph{social distancing} do not act on $\b$, but on $\a$; albeit in general $\a = \b/\ga$, in studying the effect of restrictive measures it is more convenient to consider the reduction factor $r$. That is, if the fit of the initial phase of the epidemic yields $\a_0 = \b_0 /\ga_0$ (where $\ga_0$ is determined through the formulas of Sections \ref{sec:sirfit} and \ref{sec:asirfit}), we consider in later phases a contact rate
\beql{eq:r} \a \ = \ r \ \a_0 \ , \ \ \ \ 0 < r < 1 \ . \eeq
We will see that after the first set of measures in Italy we got $r \simeq 0.5$ (compared with the initial period of the epidemic \cite{Gcov}), and after the second set of measures this went to $r \simeq 0.2$, and to $r \simeq 0.08$ at a still later stage.

{

\medskip\noindent
{\bf Remark 17.} It should be noted however that the studies suggesting the presence of a \emph{very} large fraction of undetected infections \cite{Oxf} are also implicitly suggesting that the reduction in the epidemic growth could be due not only -- or not so much -- to the containment measures, but rather to the intrinsic dynamics of the system, and on the rapid depletion of the susceptibles reservoir. This point should be carefully considered. We find that in the case of COVID-19 in Italy our model suggests $\xi = 0.1$, and discuss this point in view of that estimate. In order to fully explain the decrease of the contact rate by this mechanism -- thus essentially to the often mentioned ``herd immunity'' -- however, one would need $\xi \simeq 0.01$, which appears to be non realistic (on the other hand, the suggestion $\xi = 0.02$ \cite{Imphid} is still very low but cannot be discarded \emph{apriori}) . \EOR

\medskip\noindent
{\bf Remark 18.} The effect of different types of measures, let us say those acting on social distance and hence on $\a$ on the one hand, and those acting on early isolation and hence on $\b$ (and on $\eta$ for the A-SIR model) on the other hand, on the epidemic timescale is specially relevant in view of the economic and social costs of a generalized lockdown as the one adopted in many European countries. This is discussed in detail in related publications \cite{Cadoni,CG1,CG2}, both for the SIR and the A-SIR models. \EOR

}

\bigskip

%We thus run also a number of simulations at fixed $\b$ and varying $r$; these can give an idea of the impact of containment measures based on social distancing and hence on lowering $\a$ on the development of the epidemic within the framework of models considered here. See Figures \ref{fig:plotsir}, \ref{fig:plotasir21} and \ref{fig:plotasir14}.

However, the real concrete interest of this study is in a different point. That is, there is considerable debate on the most appropriate way to use laboratory exams, and in particular if there should be a generalized COVID testing, at least of those having had contacts with known infects, or if only clinically suspect cases should be tested. We are of course aware that the real obstacle to a generalized testing -- which should however be repeated over and over to be sure the individual has not been infected since the last test -- is of practical nature, as testing a population of several tens of million people (not to say about China or India) is unfeasible, so that this alternative is a concrete one only in small communities; these could be isolated areas or also e.g. the community of people working in a Hospital.\fofootnote{We recall anyway that when this was conducted in the small community of Vo' Euganeo, the study led to a change in our basic understanding of the COVID spreading \cite{Crisanti}.}

On the other hand, a strategy aiming at generalized testing of those who have been in contact with known infectives is more feasible; actually this strategy was conducted with remarkable success in at least one Italian region, i.e. Veneto.

In any case, we want to study what the impact of reducing $\tau = \eta^{-1}$, thus raising $\eta$, would be on the development of the A-SIR dynamics. This is illustrated in Tables \ref{tab:III} and \ref{tab:IV} below.

\subsection{More detailed study with selected parameters}
\label{sec:moredetail}

As mentioned above, our numerical simulations suggest that the epidemic in Northern Italy is -- or more precisely, was before the restrictive measures of March 8 went into operation -- better described, in terms of our model, by the situation with \beql{eq:goodpar} \b = 1/7 \ , \ \ \eta = 1/21 \ ; \ \ \xi = 1/10 \ . \eeq We will thus devote further analysis to this setting.\fofootnote{We have also conducted a detailed study of the case $\xi = 1/7$, as suggested by Li {\it et al.} \cite{Li}; but the agreement with epidemiological data is definitely less good than for $\xi= 1/10$ and hence these simulations will not be shown.}

When discussing if and how we can change these parameters, it is essential to state clearly what the two classes $I$ and $J$ (and hence also $R$ and $U$) represent when we act on the system.

What we mean here is that in the ``natural'' situations $J$ represents the class of asymptomatic \emph{and hence} undetected infectives\footnote{This was in particular true at the beginning of the COVID epidemic in Italy, when the amplitude of the asymptomatic infections was not realized; in fact,the WHO report on China was not alerting about the relevance of the phenomenon, and this was realized only after the study of the Padua group \cite{Crisanti}. See also \cite{asy1,asy4,asy8,asy9,asy10,asy13}}; on the other hand, when we start chasing for asymptomatic infectives these two characteristics are \emph{not} equivalent.

Here we will understand that $J$ represents \emph{asymptomatic} infectives, detected or undetected as they are.

It should be stressed that -- at least in this framework, see below -- these parameters cannot be altered: indeed, $\xi$ depends on the interaction of the virus with human bodies and is thus fixed by Nature, while the removal time $\b^{-1} \simeq 7$ can hardly be compressed considering that typically the first symptoms arise after 5 days, but these are usually weak and thus receive attention (especially in a difficult situation like the present one) only after some time.

{
\medskip\noindent
{\bf Remark 19.} We said above that the parameters $\b$ and $\xi$ can not be altered ``in this framework''; this requires a brief explanation. As for $\xi$ this statement is rather clear\footnote{It could depend on genetics, and thus differ in different continents, but for a given country it cannot be altered, as far as we know at the moment. Needless to say, if a medical treatment preventing the insurgence of symptom would become available, it would alter $\xi$ but at the same time it would to a great extent solve the COVID problem.}, and is connected to considering $J$ as the class of asymptomatic infectives, while if they were meant to be the class of registered infectives, $\xi$ could be altered by a large scale campaign of tests on asymptomatic population. On the other hand, for what concerns $\b$ the impossibility of compressing the time from infection to isolation is ``in this framework'' in the sense that in it isolation depends on the display of symptoms (plus some unavoidable reaction delay). The time $\de = \b^{-1}$ could instead be compressed by a campaign of \emph{tracing contacts} of known (symptomatic or asymptomatic) infectives, as done in the field at V\`o Euganeo \cite{Crisanti}. These points will not be discussed here, but will be considered in a related paper \cite{Gavsb}. \EOR
}
\bigskip

On the other hand, it is conceivable that $\tau = \eta^{-1}$ could somehow be compressed if a general screening was conducted. At the same time, the contact rate $\a$ can be reduced by a more or less rigorous lockdown; in our discussion, this reduction is encoded in the reduction parameter $r$, which yields the ratio of the achieved contact rate over the ``natural'' one -- i.e. the one measured at the beginning of the epidemic.

We have thus ran several numerical simulations for these values of $\b$ and $\xi$, both for a total population of $N = 2*10^7$ (the total population of the initially most affected regions) and for $N = 6*10^7$ (the total population of Italy).

These give of course very similar results -- if referred to the total population -- as our estimates for the parameters, and in particular for the one leading the dynamics, i.e. $\ga$, depend themselves on $S_0$.

Moreover, the questions discussed in this subsection do not concern the early phase of the epidemics (which was then limited to Northern Italy), but its future development.  We will thus present the results directly for the case $N = S_0 = 6*10^7$.

We have investigated two questions:

\begin{itemize}

\item[(A)] How a reduction in the removal time for asymptomatic infectives, i.e. in $\tau = \eta^{-1}$, would affect -- according to the A-SIR model -- the dynamics and the basic epidemiological outcomes of it in the regime where the epidemic is taking place;

\item[(B)] In the case $r$ is small enough to make the population below the epidemic threshold, what are the basic epidemiological outcomes predicted by the model, again depending on various parameters including $\eta$.

\end{itemize}

The results of these numerical investigations are summarized in Table \ref{tab:III} and Table \ref{tab:IV} respectively. We have also studied, for comparison, question (B) in the framework of the standard SIR model (question (A) can not be set in this framework). The outcomes of this study are summarized in Table \ref{tab:V}.

These Tables show that a reduction of $\tau = \eta^{-1}$ can have a significant impact -- more or less relevant depending on the $r$ parameter -- in the main epidemiological parameters, such as the infection peak, the epidemic time-span, and the fraction of the population which goes through infection with or without symptoms. (These results should be seen as preliminary, see the companion paper \cite{Gavsb} for a more detailed study.)

The point we want to stress here is that even in this concrete case \emph{the predictions of the A-SIR model}, taking into account the presence of a large set of asymptomatic infectives, \emph{differ from those of the standard SIR model}; this difference is in some cases quite significant.

{

\medskip\noindent
{\bf Remark 20.}
As mentioned above, these results are \emph{not} a forecast of the development of the COVID epidemics, as they are obtained under the hypothesis of constant parameters, while political action will drive modification of the effective parameters in one way or the other. \EOR
}
%\bigskip

%\newpage

\begin{table}
  \centering
  \begin{tabular}{||l|c||c|c||c|c|c||}
\hline
 & $r$ & $I_*$ & $t_*$ & $R_\infty / S_0$ & $U_\infty / S_0$ & $S_\infty / S_0$\\
\hline
$\xi = 1/10$  & 1.0 & $1.3*10^6$ & 57 & 0.10 & 0.89 & 0.01 \\
$\eta = 1/21$ & 0.8 & $1.0*10^6$ &  70 & 0.10 & 0.87 & 0.03 \\
 & 0.6 & $6.5*10^5$ &  95 & 0.09 & 0.82 & 0.09 \\
 & 0.4 & $2.5*10^5$ & 167 & 0.07 & 0.65 & 0.28 \\
\hline
\hline
\end{tabular}
  \caption{{  Simulations for the A-SIR model on a population of $S_0 = 6*10^7$, with $\b = 1/7$ and for the fitted initial conditions discussed in Section \ref{sec:ItaFit}, for $\xi = 1/10$ and $\eta = 1/21$, for various values of the reduction factor $r$. We report the maximum of the (registered) infectives $I_*$, the time $t_*$ at which this maximum is reached, and the fraction of the initial population which passed through the infection being registered ($R_\infty / S_0$) or unknowingly ($U_\infty/S_0$); the remaining fraction of population $S_\infty/S_0$ remains not covered by immunity.}}\label{tab:III}
\end{table}

\begin{table}
  \centering
\begin{tabular}{||l|c||c||c|c||}
\hline
 & $r$ & $t_e$ & $R_\infty / S_0$ & $U_\infty / S_0$ \\
\hline
$\xi = 1/10$  & 0.2 & 539 & $1.02*10^{-4}$ & $9.17*10^{-3}$ \\
$\eta = 1/21$ & 0.1 & 107 & $2.67*10^{-4}$ & $2.40*10^{-3}$ \\
 & 0.05 & 66 & $2.09*10^{-4}$ & $1.88*10^{-3}$ \\
 & 0.02 & 50 & $1.88*10^{-4}$ & $1.69*10^{-3}$ \\
 & 0.01 & 46 & $1.82*10^{-4}$ & $1.64*10^{-3}$ \\
\hline
\hline
\end{tabular}
    \caption{{  Simulations for the A-SIR model on a population of $S_0 = 6*10^7$, with $\b = 1/7$ and for the fitted initial conditions discussed in Section \ref{sec:ItaFit}, for $\xi = 1/10$ and $\eta = 1/21$, for various values of the reduction factor $r$ such that the population is below the epidemic threshold. We report the time $t_e$ at which there are less than 100 known infectives, and the fraction of the initial population which passed through the infection being registered ($R_\infty / S_0$) or unknowingly ($U_\infty/S_0$).}}\label{tab:IV}
\end{table}

\begin{table}
  \centering
\begin{tabular}{||l||c||c||l||c|c||}
\hline
 $r$ & $t_e$ & $R_\infty / S_0$ &  $r$ & $t_e$ & $R_\infty / S_0$ \\
\hline
0.20 & 57 & $2.29*10^{-4}$ & 0.20 & 41 & $1.77*10^{-4}$ \\
0.10 & 49 & $1.98*10^{-4}$ & 0.10 & 37 & $1.58*10^{-4}$ \\
0.05 & 46 & $1.87*10^{-4}$ & 0.05 & 35 & $1.51*10^{-4}$ \\
0.02 & 44 & $1.81*10^{-4}$ & 0.02 & 34 & $1.47*10^{-4}$ \\
0.01 & 44 & $1.79*10^{-4}$ & 0.01 & 34 & $1.46*10^{-4}$ \\
\hline
\hline
\end{tabular}
    \caption{{  Simulations for the standard SIR model on a population of $S_0 = 6*10^7$, with $\b = 1/7$ (left hand side) and -- for comparison -- also for $\b = 1/5$ (right hand side), and for the fitted initial conditions discussed in Section \ref{sec:ItaFit}, for various values of the reduction factor $r$ such that the population is below the epidemic threshold. We report the time $t_e$ at which there are less than 100 known infectives, and the fraction of the initial population which passed through the infection ($R_\infty / S_0$).}}\label{tab:V}
\end{table}

\section{COVID-19 in Italy and mitigation measures}
\label{sec:measures}

Our study in the previous Sections went until March 17 (see Figs.\ref{fig:plotsirbest} and \ref{fig:plotasirbest}); this is the time interval in which it makes sense to consider \emph{constant} parameters, as the first set of governmental measures did not show its effect yet, and we got a good fit of epidemiological data by our A-SIR model.

\subsection{Mitigation measures and reduction factors}

After (or shortly before) this date, the first set of restrictive measures, taken on March 8 and gradually enforced, should have shown their effect; thus -- in terms of the model -- the parameter $\a = \a_0$ changed to a different value $\a = \a_1  = r_1 \a_0$. Similarly, more restrictive measures took effect on March 23 (this time in a sharper way), and this should have shown their effect about one week later, thus changing again the parameter $\a$, say taking it to be $\a = \a_2 = r_2 \a_0$. It should be mentioned that albeit no official change took place, it was generally remarked that a different attitude in enforcing the measures appeared after Easter (April 11); at about the same time individual protective device became easily available. These changes are somehow reflected in our fitting, see below, by the introduction of a different reduction parameter $r_3$ after April 25.

Thus in considering if our model can describe the actual development of the epidemic in Italy, one should take into account these changes of parameters, i.e. fit also the constants $r_1$, $r_2$ and $r_3$ introduced above. It should be stressed that we do not have an analytical formula involving some parameters which can then be fitted against experimental data; we are instead studying numerical solution of the A-SIR dynamical system \eqref{eq:ASIR} and checking how this fits the data. On the other hand, we are assuming that nothing changes for the $\b$, $\eta$ and $\xi$ parameters.

Data until May 15 are given in Table \ref{tab:I}.
We found a good agreement keeping the value $\xi = 1/10$, and setting
\beql{eq:rmeas} r_1 \ = \ 0.50 \ , \ \ \ r_2 \ = \ 0.20 \ , \ \ r_3 \ = \ 0.08 \ . \eeq
In other words, we have set (recall time is measured in days, day 1 being February 21)
\beql{eq:ameas} \a \ = \ \a (t) \ = \ \cases{ 1.00 * \a_0 & $t \ \le \ 25$ \ , \cr
0.50 * \a_0 & $25 < t \le 35$ \ , \cr
0.20 * \a_0 &  $35 <  t \le 63$ \ , \cr
0.08 * \a_0 & $63 < t$ \ . \cr}  \eeq

{

\medskip\noindent
{\bf Remark 21.}
Not that here we are considering the full Italian population, i.e. $S_0 = 6*10^7$; this introduces a factor 3 compared with the setting used in Sections \ref{sec:Italy} and \ref{sec:numerical} to fit $\a_0$ on the basis of the early data which concerned essentially three regions with a total population of about $2*10^7$; we have correspondingly to divide the $\a_0$ determined in there by a factor three, thus getting the value \beq \a_0 \ = \ 3.77*10^{-9} \ ; \eeq
this is to be used in \eqref{eq:ameas} above. \EOR
}
\bigskip

\begin{figure}
\centering
  \includegraphics[width=200pt]{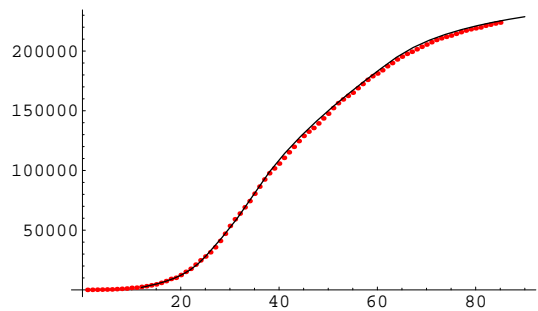}\\
  \caption{{  Solution of the A-SIR equation, taking into account the changes in the contact rate $\a$ determined by governmental measures, against epidemiological data for Italy. See text.}}\label{fig:measures}
\end{figure}

\begin{figure}
\centering
  \includegraphics[width=100pt]{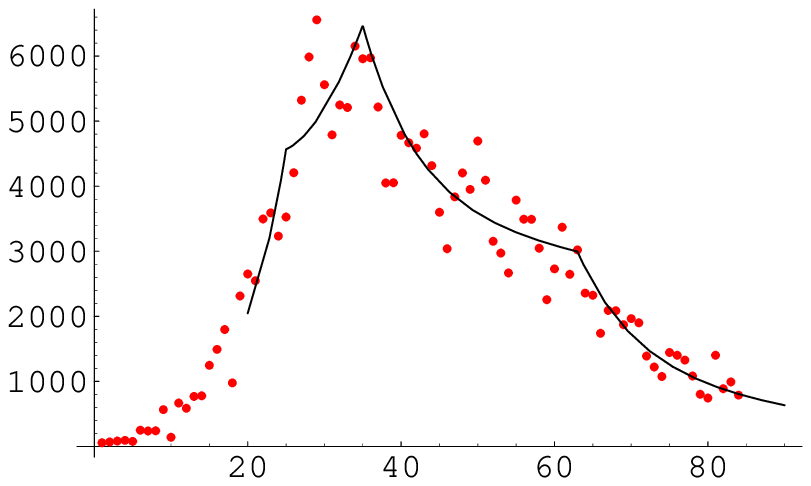} \ \
  \includegraphics[width=100pt]{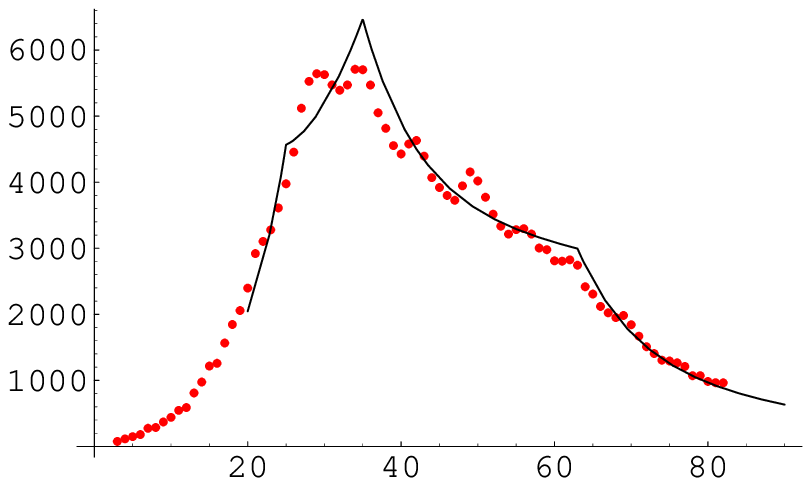} \\
  \caption{{  Data for $I(t)$ as obtained from the model equation, i.e. as $\b I(t) = R' (t)$, against data for the $R(t)$ increment over one day. Left: raw data; Right: data smoothed by averaging over five days, centered at plotted day. We are actually plotting daily increments of $R$, i.e. the A-SIR estimate for $\b I(t)$.}}\label{fig:RP}
\end{figure}

Proceeding in this way, we obtain the curve plotted in Fig.\ref{fig:measures} against the epidemiological data; this curve shows a good agreement with the data. See also Fig.\ref{fig:RP} for a different representation in terms of the infectives $I(t)$ (in one of the panels we have adopted a smoothing procedure consisting in mobile average over five days).

\newpage

\subsection{COVID-19, Italy, and mitigation measures: a brief discussion}

We give here a brief discussion of several aspects of the application of our theory to the specific situation of the COVID-19 epidemics in Italy, and of the results obtained above. This also involve non-mathematical matters. Moreover, the points touched upon here are not applying to the general theory, so we have preferred to keep this discussion separate from the short general one given in the next Section \ref{sec:conclu}.

First of all, we note that the model allows to give an estimate also for the lockdown duration (assuming no further measures or relaxation of measures is adopted in the meanwhile, which is of course improbable in practical terms). E.g., if the lockdown should go on until the level of registered infectives present when the first measures were adopted on day 17, i.e. $I(t) \simeq 7,000$, then according to the model and disregarding the third reduction, it should have gone on until day 129, i.e. the end of June. Taking into account also the third reduction in $\a$, it should have gone until day 79, so May 19, in quite good agreement with the actual choice of the Italian Government to reduce limitations starting on May 18. We stress that according to our model and more generally to SIR-type models, a strategy based also on \emph{tracing contacts} and \emph{early detection} is much more effective (also in stopping the epidemic, as shown in the field by the Padua team \cite{Crisanti}, besides in keeping restrictions within an affordable time), as discussed in a companion paper \cite{Gavsb}; see also \cite{Cadoni,CG1,CG2}.

\begin{figure}
\centering
  \includegraphics[width=100pt]{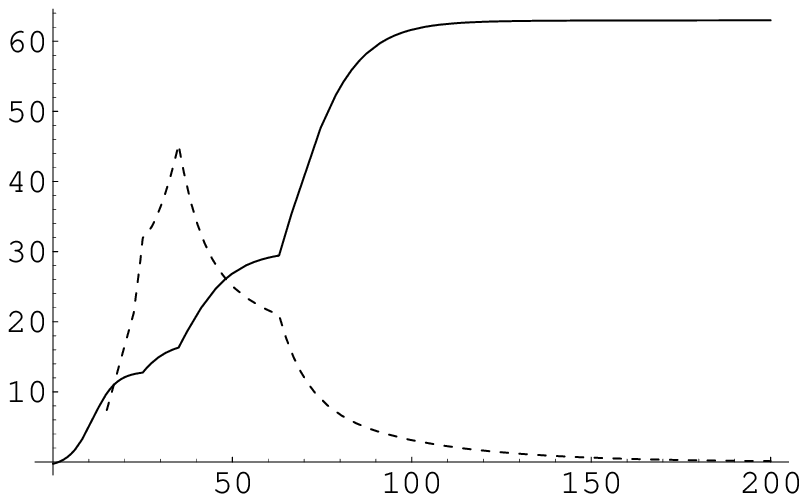} \ \
  \includegraphics[width=100pt]{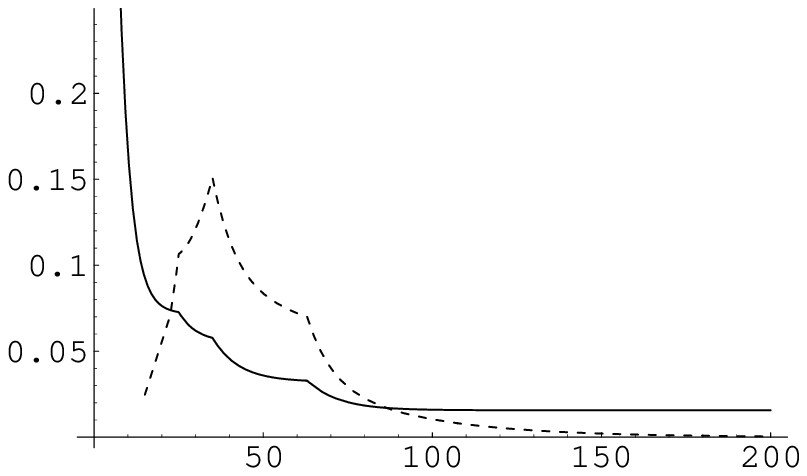} \\
  \caption{{  The balance between symptomatic and asymptomatic infectives. Left: the ratio $J(t)/I(t)$ (solid) and $I(t)$ (dashed, different scale) on the numerical solution to A-SIR equations. Right: the ratio $x(t) = I(t)/[I(t)+J(t)]$ (solid) and $I(t)$ (dashed, different scale) on the numerical solution to A-SIR equations.}}\label{fig:ratio}
\end{figure}

Estimating the number of asymptomatic infectives is of course relevant in choosing measures to counteract the COVID-19 epidemics; as far as we know this is the first estimate based on a theoretical model and not just on statistical considerations. It follows from our model that the fraction $x$ of symptomatic infectives -- and the fraction $y = 1-x$ of asymptomatic ones -- are \emph{dynamical} variables, and change with time depending on other features of the system, in particular its total population. Thus a purely statistical evaluation appears to be \emph{necessarily misleading}. See also Fig.\ref{fig:ratio} in this regard.

The previous observation is specially significant when we look at the tail of the epidemic. In fact, while in this limit the ratio of removed infectives $U(t)/R(t)$ goes to the natural limit $(1-\xi)/\xi$, the ratio of \emph{active} infectives $J(t)/I(t)$ goes to a much larger number, see Fig.\ref{fig:ratio}. This means that when restrictions are removed, it is absolutely essential to be able to track asymptomatic infectives, to avoid these can spark new fires of infection.

It should also be mentioned that our analysis suggests that the restrictive measures adopted in Italy were quite successful in reducing the contact  rate $\a$. This is in a way surprising, given their mild nature if compared to those adopted in China. A couple of facts should be recalled in this respect: $(a)$ on the one hand a large part (possibly larger than in other European countries with a similar demographic profile) of infections and casualties is related to hospitals or senior citizens residences, so that once these were put under control the transmission rate was substantially lowered;  the latter are much more common in the Northern regions; $(b)$ a reduction on $\a$ has an impact on the (only) quadratic term in the SIR or A-SIR equations; thus a reduction in mobility by a factor $\mu$ could result in a reduction by a factor $\mu^2$ for the contact rate $\a$.\footnote{In this respect, it should be mentioned that Google has provided data analyzing the reduction of mobility in different contexts for different countries \cite{google}. These are difficult to compare across countries, as the restrictions have not been the same in different countries; there is however a homogeneous sector, as in all countries the access to food and medicine shops remained unrestricted. In these sector, according to the Google data for March, Italy is -- among the large European countries severely affected by COVID -- the one where the reduction in this sector was more marked. We have indeed  -85\% for Italy, -76\% for Spain, -62\% for France, -51\% for Germany, - 46\% for Great Britain.
Thus the reduction of $\a$ could be not only, or not so much, due to the governmental measures, but to the more prudent attitude of citizens.} \EOR

Finally, we note that while the first two steps in the $\a$ reduction are temporally correlated to the restrictive measures adopted by the Government and are thus thought to be due to these, the third step cannot be explained in such a way; several tentative explanations can be put forward (but with available data not scientifically tested) for this. These include relaxation of ``stay at home'' campaign (home is second only to senior citizens residence for frequency of contagion), availability and thus generalized use of individual protection device, increased solar UV radiation, and reduction of ambient viral load \cite{Volpert}. No final word on this matter can of course be said in terms of mathematical models alone.

\section{Conclusions}
\label{sec:conclu}

Motivated by the peculiar features of the COVID epidemics, we have considered a SIR-type model, called A-SIR model, taking into account the presence of a large set of \emph{asymptomatic infectives}.

We have shown that the dynamics of the SIR and the A-SIR models, for parameters fitted from the \emph{same set of data} available in the early phase of an epidemic, differ significantly; this is not surprising, but is in itself a significant conclusion when we have to deal with a concrete epidemics with this characteristic. This is our first result.

We have analyzed the available data for the COVID-19 epidemics in Northern Italy in terms of the SIR and of the A-SIR models; in particular we have fitted the model parameters based on the period 1-10 March, and considered how these models with such parameters are performing in predicting the evolution for the subsequent week, 11-17 March. As shown by Figure \ref{fig:plotsirbest} on the one hand, and by Figure \ref{fig:plotasirbest} on the other hand, it appears that the A-SIR model is much better in predicting such (admittedly short time) evolution. In particular, this is the case with the estimate $\xi =1/10$ for the ratio of clearly symptomatic versus total infections (this is slightly smaller that the Li {\it et al.} \cite{Li} estimate $\xi = 1/7$); and with the reasonable estimates $\b^{-1} = 7$ days for the time from infection to isolation for symptomatic infectives, and $\eta^{-1} = 21$ for the time from infection to healing of asymptomatic infectives.

We have studied in more detail the case which best fits the epidemiological data outside the period used to fix the model parameters. This study was conducted in two directions.

\medskip\noindent
{\bf (A)} On the one hand, we have considered what would be the effect of a reduction of the time spent by asymptomatic being infective and non-isolated. In this framework, two cases are possible: either the restrictive measures are only mitigating the epidemic, or they are capable of stopping it by raising the epidemic threshold above the population level. In the first case, a reduction of $\eta^{-1}$ from 21 to 14 days produce a substantial lowering of the epidemic peak and also substantially postpones its occurrence; in the second case, the effect of such a reduction may be quite relevant if the population remains just under the threshold (see the case with $r=0.2$ in Tables \ref{tab:IV}), or not so relevant if the reduction of the contact rate is taking the epidemic threshold well above the population level (see the cases with lower $r$ in Table \ref{tab:IV}).

In all cases, \emph{there is a marked difference with the behavior of a standard SIR model with equivalent parameters}.

\medskip\noindent
{\bf (B)} On the other hand, we have then considered how our model can describe the COVID-19 epidemic dynamic in Italy outside the time window used to fit the parameters. In this context, restrictive measures adopted in different stages have reduced the epidemic development, hence also altered the parameters -- in particular, as the measures were essentially based on \emph{social distancing}, the contact rate $\a$. We have seen that assuming the measures showed their effect after one week (in line with our estimate $\b^{-1} \simeq 7$) there is an estimate of their effect on the contact rate $\a$ which produces \emph{a good agreement between the model and the data}.

This agreement depends on the chosen value for $\xi$, i.e. the ratio between symptomatic infections and total infections, and is quite good for $\xi \simeq 1/10$ (confirming the early estimate by Vallance and Whitty \cite{bbc}). It was stressed that this estimate of $\xi$ is not of statistical nature, but follows instead from a theoretical model; this is specially relevant in that the same model shows that albeit the probability of symptomatic infection $\xi$ is a constant, the different times for which symptomatic and asymptomatic infectives do take part in the epidemic dynamic makes that the fraction $x$ of symptomatic over total infectives is a dynamical variable, and changes -- even substantially -- with time.\footnote{In this respect, it should be mentioned that recently several Groups have suggested the fraction of detected infectives could be, for COVID, even smaller \cite{Oxf,Imphid}; this would by all means make even greater the differences between the dynamic predictions by a standard SIR model or by a model, like the simple A-SIR model we propose here, taking into account the peculiar feature of the presence of a large class of asymptomatic infectives.}
\bigskip

We trust that our work shows convincingly the need to take into account the presence of asymptomatic infectives -- and the longer time they spend before going out of the infective dynamics -- when they are a substantial number; see also the Appendix and \cite{GR0}.

The ongoing COVID-19 epidemic taught us that there can be relevant epidemics with a large number of asymptomatic infectives; more detailed models can surely be cast, but the very simple model presented here is already sufficient to show that the standard SIR model is not a good guide in this case, as it leads to overestimate certain very relevant parameters and underestimate others.

A more detailed study of how different strategies to mitigate the epidemic may affect the A-SIR dynamics is presented in a related paper \cite{Gavsb}; see also \cite{CG1,CG2}.

\section*{Acknowledgements}

\noindent
I thank L. Peliti (SMRI), M. Cadoni (Cagliari) and E. Franco (Roma) for useful discussions. The paper was prepared over a (locked-in) stay at SMRI. The author is also a member of GNFM-INdAM.

\begin{appendix}

\section{A-SIR versus SIR dynamics, and $R_0$ estimates}

The \emph{basic reproduction number} (BRN) is a popular indicator for the speed of diffusion of an epidemic in its first stages \cite{Diek,Diet,Hee}. This is usually denoted as $R_0$, but which here will be denoted as $\rho_0$ (to avoid any confusion with the value of $R(t)$ at $t=0$).

The BRN represents the average number of infections caused by a single infective; we want to discuss how the estimate of this can be flawed if one works with the SIR formalism in a situation where a large number of asymptomatic infectives is present, and hence better described by the A-SIR model; our discussion reproduces that in \cite{GR0}, and is given here for the sake of completeness.

\subsection{The reproduction number and the SIR framework}

{ The extraction of the BRN from epidemiological data can be tricky, but its evaluation in terms of a model is much simpler. In fact, in this case $S(t)$ can be considered, for small $t$, as constant, i.e. $S(t) \approx S_0$. The SIR equations reduce to linear ones, and in the very beginning (before any infective can be removed) the new infectives grow as $dI/dt = \a S_0 I$. Thus any infective produces $\a S_0(\de t)$ new infectives in a short time length $\de t$. As each infective is active, on the average, for a time $\b^{-1}$, this yields a first rough estimate
\beql{eq:R00} \rho_0 \ \simeq \ (\a/\b) \, S_0 \ = \ S_0 / \ga \ . \eeq
Obviously, as $\b^{-1}$ is not so small (of the order of days), we should look more carefully at the equation for $I(t)$, which can be solved and yields
$I(t) = \exp \[  \a S_0 t \] I(0)$.
As each infective is active for an average time $\de = \b^{-1}$, this just reads
\beql{eq:R01} I(\de ) \ = \ \exp \[ S_0/ga \] \ I(0) \ . \eeq
The situation is slightly different if we take into account also the removal of infectives. With the same approximation $S \approx S_0$ but with the full \eqref{eq:SIR} equations, we get $dI/dt = \a (S_0 - \ga) I$ and hence
$$ I(t) \ \approx \ \exp [ \a (S_0 -\ga) \, t ] \ I_0 \ := \ H(t) \, I_0 \ . $$
Considering again that an infective remains active for a time $\de = \b^{-1}$ on average, the previous equation means that (on average) there will be $\rho_0 = H(\de)$ new infectives originating from a single one. From the previous expression for $H(t)$ we easily get
$ \rho_0 = \exp [ (S_0 / \ga) - 1 ]$.

If we are facing a pathogen for which  there is no natural immunity in the population (e.g. a new virus), $S_0 = N$ (the total population) and we get
\beql{eq:R02} \rho_0 \ = \ \exp \[ \frac{N}{\ga} \ - \ 1 \] \ . \eeq
Thus, evaluating $R_0$ is in this case immediate if we know $\ga$; and conversely if we are able to evaluate $\rho_0$ by epidemiological data, then  it is immediate to evaluate $\ga$ as \beql{eq:R03} \ga \ = \ \frac{N}{1 + \log (\rho_0) } \eeq
(or a simpler formula if we adopt the earlier and simpler definitions for $\rho_0$, e.g. \eqref{eq:R00} above).

The parameter $\rho_0$, and more generally $\rho (t)$, met a great favor in the general press, as it gives the impression to describe a complex phenomenon by a single number. As we mentioned above, evaluating $\rho (t)$ from epidemiological data can be a tricky matter \cite{Diek,Diet}, in particular if this is attempted without resorting to a specific model.

In this case too, using the SIR model without taking into account the presence of a large set of asymptomatic infectives leads to a wrong estimation of $\rho$, as we will see in a moment. }

\subsection{The reproduction number and the A-SIR framework}

With the A-SIR equations \eqref{eq:ASIR}, the number of new infected per unit of time is $\a S (I+J)$, and in the early phase of the epidemic we can assume $S \approx S_0$, and infectives will grow as
$$ K(t + \de t) \ = \ K(t) \ \exp[\a \ S_0 \ \de t] \ . $$ Thus each infective will give origin in the time span $\de t$ to $\exp[\a S_0 \de t] $ new infectives. In order to know the BRN, we we should look at the \emph{average infective time} $\tau$ in the early phase of the epidemic, and choose $\de t = \tau$. At the beginning of the epidemic the ratio between registered and total infectives is simply
\beq x_0 \ := \ \frac{I_0}{I_0+J_0} \ \simeq \ \xi \ , \eeq
while -- as discussed above -- in later stages the proportion between $I$ and $J$ changes, as individuals stay longer in the $J$ class than in the $I$ class.

The average removal rate for $t \simeq 0$ is thus
\beql{eq:B} B_0 \ = \ \xi \, \b \ + \ (1-\xi) \, \eta \ . \eeq

This means that each (symptomatic or asymptomatic) infective individual will give direct origin, across its infective and non-isolation period, not to $\rho_0 = \exp[ \a S_0 / \b] = \exp [S_0/\ga]$ new infectives, but instead to
\beql{eq:R0h} \^\rho_0 \ = \ \exp[(\a/B) S_0] \ = \ \( \rho_0 \)^{\b/B} \eeq
new infectives. As $\b > B$, this means that the actual basic reproduction number $\^\rho_0$ is larger -- and possibly substantially larger -- than the value $\rho_0$ which would be estimated solely on the basis of registered infections.

A trivial computation on the basis of the values given above -- i.e. $\b^{-1} \simeq 7$, $\eta^{-1} \simeq 21$, $\xi \simeq 1/10$ -- provides
\beql{eq:R0corr} \^\rho_0 \ = \ \( \rho_0 \)^{5/2} \ . \eeq
Recalling that the estimates of the COVID $\rho_0$ on the basis of registered infectives are in the range $\rho_0 \in (2.5 - 3.0)$, this means that the actual BRN turns out to be instead in the range
\beql{eq:R0A1} \^\rho_0 \ \in \ \( 10 - 15 \) \ . \eeq

This could explain why all Health Systems were surprised by the rapid growth of the number of COVID-19 infections; in fact, the presence of a large set of asymptomatic infectives was not realized when the epidemic attacked the first countries, and is becoming clearly established only now.

{

\medskip\noindent
{\bf Remark A.1.} In terms of the standard SIR model, our discussion shows that one should consider $B_0$ -- see eq.\eqref{eq:B} -- rather than $\b$ in computing the BRN. This also explains the poor performance of the SIR model in fitting Italian epidemiological data with ``medically reasonable'' estimates for $\b$; in fact, considering the $B_0$ value instead of $\b$ produces a good enough fit \cite{CG2}. { See also Remark 12 in this context.} \EOR

\medskip\noindent
{\bf Remark A.2.} On the other hand, it should be noted that according to Britton, Ball and Trapman \cite{BRT} the herd immunity level for COVID-19 is substantially lower than the classical herd immunity level; this yields a nicer perspective for the future. \EOR
}

{
\medskip\noindent
{\bf Remark A.3.} Note that eq.\eqref{eq:R0corr} is obtained considering the A-SIR equivalent of eq.\eqref{eq:R03}. Had we stayed with the rough estimate \eqref{eq:R00} (as in \cite{GR0}), its A-SIR equivalent would have given
$\^\rho_0  =  (5/2) \rho_0$,
which means that the actual BRN is in the range
$\^\rho_0 \in ( 6.25 - 7.5 )$, i.e. still substantially higher that the SIR estimation of $\rho_0$. \EOR
}

\end{appendix}

%\newpage

\end{document}